\documentclass[11pt,a4paper,aps,preprint,superscriptaddress,nofootinbib]{revtex4-1}
\usepackage{graphicx}
\usepackage{amssymb}
\usepackage{amsmath}
\usepackage{epstopdf}
\usepackage{color}
\usepackage[normalem]{ulem}
\usepackage[colorlinks=true, pdfstartview=FitV, linkcolor=blue, citecolor=blue, urlcolor=blue]{hyperref}

\begin{document}

\title{Sensitivity of future lepton colliders to the search for charged lepton flavor violation}

\author{Tong Li}
\email{litong@nankai.edu.cn}
\affiliation{
School of Physics, Nankai University, Tianjin 300071, China
}
\author{Michael A.~Schmidt}
\email{m.schmidt@unsw.edu.au}
\affiliation{
School of Physics, The University of New South Wales, Sydney, New South Wales 2052, Australia
}



\begin{abstract}
The observation of lepton flavor violation indicates new physics beyond
the Standard Model. Lepton colliders are ideal facilities to probe charged
lepton flavor violation (CLFV) signals induced by new physics at high energy. In
this work we perform a comprehensive study of the sensitivity of future lepton
colliders to charged lepton flavor violation. We consider the most general renormalizable
Lagrangian coupling two leptons to new bosonic particles,
involving both $\Delta L=0$ and $\Delta L=2$ interactions. The CLFV processes
are introduced by the exchange of off-shell new particles at tree-level. We find that CEPC, ILC, FCC-ee, and CLIC each provide a complementary probe of CLFV couplings to low-energy precision experiments for $\tau$ lepton(s) in final states, while low-energy precision experiments are more sensitive in the absence of $\tau$ leptons.
\end{abstract}


\maketitle

\section{Introduction}
\label{sec:Intro}

The observation of lepton flavor violation (LFV) clearly indicates new physics beyond the Standard Model (SM).
In the SM, lepton flavor numbers are conserved because neutrinos are massless.
The observation of neutrino oscillations and thus neutrino masses, however,
confirms the existence of LFV in the neutrino sector. Thus LFV
naturally occurs among charged leptons due to SU$(2)_L$ symmetry, that is
the charged lepton flavor violation
(CLFV)~\cite{deGouvea:2013zba,Bernstein:2013hba}.
It is defined as processes conserving total lepton number $L\equiv
L_e+L_\mu+L_\tau$ (and baryon number $B$) but violating the global symmetry group
\begin{eqnarray}
{\rm U}(1)_{L_\mu - L_\tau}\times {\rm U}(1)_{L_\mu + L_\tau - 2L_e},
\end{eqnarray}
without involving neutrinos~\cite{Heeck:2016xwg}.
Massive neutrinos themselves can lead to CLFV processes from right-handed Dirac neutrinos in the SM or from Majorana neutrinos induced by the so-called Weinberg operator. However, the induced CLFV processes are currently unobservable because they are suppressed by the unitarity of the leptonic mixing matrix and the smallness of neutrino masses. Many other neutrino mass models, such as the inverse seesaw model~(see e.g.~\cite{Malinsky:2009df}) and radiative neutrino mass models~\cite{Cai:2017jrq}, predict sizable CLFV processes. CLFV can also arise in many other extensions of the SM such as multi-Higgs doublet models or the minimal supersymmetric standard model via gauginos-slepton loops with off-diagonal terms in the slepton soft mass matrix~\cite{Raidal:2008jk}.

The search for CLFV and the corresponding new physics can be traced back to
the 1940s and 50s when the first bounds were obtained for $\mu\to
e\gamma$~\cite{Hincks:1948vr} and $\mu$-$e$ conversion in
nuclei~\cite{Steinberger:1955hfk}. CLFV has yet to be observed experimentally. Thus upper limits have been derived on the scale of CLFV of the order of 100 (10) TeV for muon (tau) decays~\cite{Calibbi:2017uvl}. There are also several scenarios of CLFV search at colliders. The ATLAS and CMS experiments searched for CLFV from Higgs decays, i.e. $H\to\ell\tau$~\cite{Aad:2016blu,Sirunyan:2017xzt}, and derived upper bounds on the off-diagonal Yukawa couplings of the order of $10^{-3}$. The ATLAS experiment currently provides the most stringent limit for $Z\to e\mu$~\cite{Aad:2014bca} with ${\rm BR}<7.5\times 10^{-7}$, while experiments at the Large Electron Positron (LEP) collider set the most stringent limits for $Z\to e\tau$ ($Z\to \mu\tau$) with ${\rm BR}<9.8\times 10^{-6}$~\cite{Akers:1995gz} (${\rm BR}<1.2\times 10^{-5}$~\cite{Abreu:1996mj}). Lepton colliders with their clean environment and well-understood backgrounds can outperform hadron colliders with less integrated luminosity and thus provide an ideal facility to probe rare CLFV events.

The proposed lepton colliders, in terms of
the center of mass (c.m.) energy and the integrated luminosity used in our
analysis, are
\begin{itemize}
\item Circular Electron Positron Collider (CEPC): 5 ab$^{-1}$ at 240 GeV~\cite{cepc},
\item Future Circular Collider (FCC)-ee: 16 ab$^{-1}$ at 240 GeV~\cite{FCCee:2017},
\item International Linear Collider (ILC): 4 ab$^{-1}$ at 500 GeV~\cite{Barklow:2015tja},
\item Compact Linear Collider (CLIC): 5 ab$^{-1}$ at 3 TeV~\cite{Charles:2018vfv}.
\end{itemize}
With their large foreseen c.m.~energies and luminosities, the search for CLFV
at lepton colliders should yield complementary results to those of searches for rare lepton
decays. In this work, we illustrate the projected sensitivity of future lepton
colliders to the CLFV signal arisen from new physics beyond the SM. Note that we will not include the results of ILC with 2 ab$^{-1}$ luminosity at 250 GeV, as the c.m.~energy of proposed ILC is very close to that of CEPC and FCC-ee and its integrated luminosity is much smaller as a linear collider. The above ILC case is stated as ILC500 in the following context and figures.
We will also consider the option of colliding two electron beams for each of the four proposed lepton colliders, assuming the same center of mass energies but a reduced integrated luminosity of 500 fb$^{-1}$.

Previous searches for CLFV at lepton colliders mainly focused on the rare decays of
a tau lepton~\cite{Aubert:2009ag}, $Z$
boson~\cite{Abada:2014cca,Antusch:2016ejd} and Higgs
boson~\cite{Arganda:2014dta,Banerjee:2016foh,Qin:2017aju}. We instead study the
scattering CLFV processes mediated by the new particles beyond the SM at lepton
colliders. The advantage of this scenario of off-shell channels is that one can
directly compare the projected limit of CLFV couplings with current low-energy
precision constraints. To apply the results to a broader extent, in the aspect of new physics contribution, we consider the most
generic interactions that are allowed by Lorentz invariance and gauge invariance.

The paper is outlined as follows. In Sec.~\ref{sec:Lag}, we describe the general SM extensions with CLFV couplings. Then, we discuss
the low-energy precision constraints on the CLFV couplings in
Sec.~\ref{sec:cons}. We present the sensitivity of future lepton colliders to
the CLFV couplings in Sec.~\ref{sec:sen}, where we also show the comparison to
the low-energy constraints. Our conclusions are drawn in Sec.~\ref{sec:Con}. Some technical details for low-energy precision constraints are collected in the Appendix.

\section{General Lagrangian for charged lepton flavor violation}
\label{sec:Lag}
In this section we construct the most general Lagrangian which couples two charged leptons to either a scalar or a vector boson. We consider both lepton number conserving bilinears $\bar\ell \Gamma \ell^\prime$ ($\Delta L=0$) and lepton number violating bilinears $\bar{\ell^c} \Gamma \ell^\prime$ ($\Delta L=2$), where $\Gamma$ denotes different $\gamma$-matrix structures. The most general Lagrangians for $\Delta L=0$ and $\Delta L=2$ bilinears are given by
	\begin{eqnarray}
	\mathcal{L}_{\Delta L=0}&=&y_1^{ij} H_{1\mu}^0 \bar{\ell}_i \gamma^\mu P_L \ell_j + y_1^{\prime ij} H_{1\mu}^{\prime 0} \bar{\ell}_i \gamma^\mu P_R \ell_j + \left( y_2^{ij} H_{2}^0 \bar{\ell}_i P_R \ell_j + H.c. \right)\;,\label{eq:delL0} \\
		\mathcal{L}_{\Delta L=2}&=&\left( \lambda_1^{ij} \Delta_1^{++}\bar{\ell^c_i}P_R\ell_j + H.c.\right) + \left( \lambda_2^{ij} \Delta_{2\mu}^{++}\bar{\ell^c_i}\gamma^\mu P_R\ell_j + H.c.\right) - \left( \lambda_3^{ij}\Delta_3^{++}\bar{\ell^c_i}P_L\ell_j + H.c. \right).\label{eq:delL2}
\end{eqnarray}
The subscript of the new bosonic fields, i.e.~1, 2 or 3, indicates the SU$(2)_L$ representation, singlet, doublet or triplet, respectively.
The terms in the Lagrangians are obtained by expanding the most general SM gauge invariant Lagrangian in terms of explicit leptonic fields~\cite{Cuypers:1996ia}.

In the $\Delta L=0$ Lagrangian, $H_{1\mu}^{(\prime)0}$ are real neutral gauge bosons, while $H_2^0=(h_2+i a_2)/\sqrt{2}$ is a complex neutral scalar. We thus have to consider both scalar and pseudoscalar parts of $H_2^0$. The couplings $y_1^{(\prime)}$ may arise from new gauge interactions with a LFV $Z'$, and $y_2$ can naturally appear in two Higgs doublet models.
Note that the interaction with an SU$(2)_L$ triplet gauge boson, i.e. $\bar{L}\gamma^\mu P_L L H_{3\mu}$, gives the same charged lepton interaction as $H^0_{1\mu}$ and thus we do not list it separately.

The $\Delta L=2$ coupling $\lambda_1$ may originate from the SU$(2)_L$ singlet field in the Zee-Babu model which only couples to right-handed charged leptons~\cite{Zee:1985id,Babu:1988ki} and $\lambda_3$ may come from the SU$(2)_L$ triplet field in the Type II Seesaw model which only interacts with left-handed charged leptons~\cite{Magg:1980ut,Schechter:1980gr,Lazarides:1980nt,Wetterich:1981bx,Mohapatra:1980yp}. The coupling $\lambda_2$ can arise after the breaking of a unified gauge model where the lepton doublet $L$ and charge-conjugate of the charged lepton singlet $\ell^c$ reside in the same multiplet. One example is an SU$(3)_c\times$SU$(3)_L\times$U$(1)_Y$ model~\cite{Pisano:1991ee}.

Two examples of ultraviolet(UV)-complete models are a LFV two Higgs doublet model whose leptonic Yukawa term is described by
\begin{align}
y_{2}^{ij} H_{2\alpha} \bar{L}_{i\alpha} P_R \ell_j  + h.c. \;,
\end{align}
and the Type II Seesaw model
\begin{align}
- \lambda_3^{ij} \Delta_{3\gamma\beta}\epsilon_{\alpha\gamma}  L^T_{i\alpha} C P_L L_{j\beta} + h.c. \;,
\end{align}
with the charge conjugation matrix $C=i\gamma_2\gamma_0$.
Besides the charged lepton terms in Eqs.~(\ref{eq:delL0}) and (\ref{eq:delL2}), as shown below, there are additional contributions to low-energy precision constraints in the above UV-complete models, which we provide in the Appendix. In the rest of this paper, however, we will restrict ourselves to the Lagrangians in Eqs.~\eqref{eq:delL0} and \eqref{eq:delL2}.

\section{Constraints}
\label{sec:cons}

We discuss three different classes of constraints on the CLFV parameters in the models introduced in the previous section. In the analysis we restrict ourselves to one new boson at a time. The first subsection discusses low-energy precision constraints from tree-level rare decays, radiative decays, the anomalous electromagnetic moments of leptons, and muonium antimuonium conversion. The second one presents existing constraints from the DELPHI experiment at the Large Electron Positron (LEP) collider and searches at the Large Hadron Collider (LHC), while the third subsection briefly describes any additional constraints which arise from embedding the particles in complete SM multiplets.

\subsection{Low-energy precision constraints}
\label{sec:lowEcons}

\begin{figure}[tpb!]
\begin{center}
\includegraphics[scale=1,width=12cm]{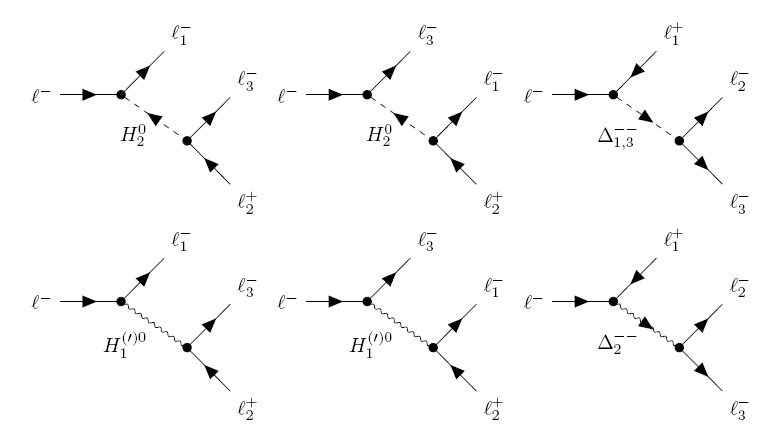}
\end{center}
\caption{Tree-level LFV trilepton decays.}
\label{LFV:tree}
\end{figure}

\begin{figure}[tbp!]
\begin{center}
\includegraphics[scale=1,width=12cm]{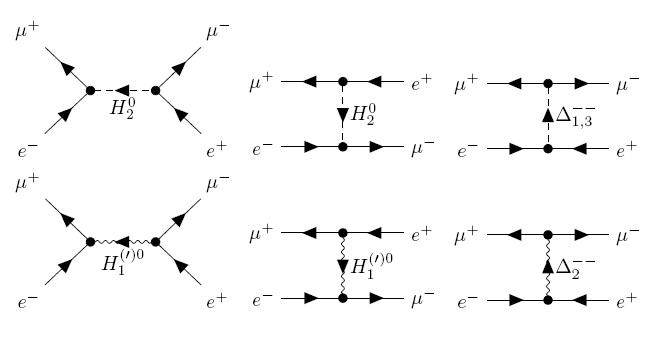}
\end{center}
\caption{Muonium-antimuonium conversion.}
\label{LFV:conversion}
\end{figure}

\begin{figure}[tbp!]
\begin{center}
\includegraphics[scale=1,width=12cm]{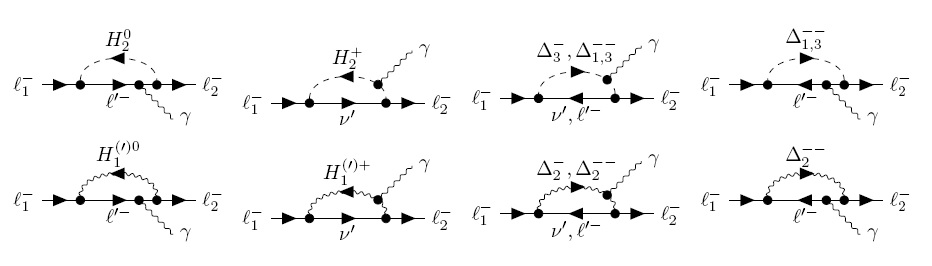}
\end{center}
\caption{Radiative LFV decays and anomalous magnetic moments.}
\label{LFV:radiative}
\end{figure}
The new bosons contribute to low-energy precision observables at either tree-level or loop-level. In particular, there are tree-level contributions to trilepton decays and muonium-antimuonium conversion, as shown in Figs.~\ref{LFV:tree} and \ref{LFV:conversion} respectively. At the one-loop level, there are radiative lepton decays from Fig.~\ref{LFV:radiative} and new contributions to electromagnetic dipole moments. Electric dipole moments are CP violating and thus are not induced by real couplings. We find that magnetic dipole moments only yield mild constraints, as long as they are not enhanced by the $\tau$ lepton mass in the loop.

The relevant constraints on the low-energy precision LFV processes are summarized in Table~\ref{tab:LE} and the theoretical predictions for the different processes are collected in the Appendix.
\begin{table}[btp!]\centering
	\begin{tabular}{|c| c|c |c|}\hline
		process & experimental constraint & future sensitivity &lepton flavor symmetry \\\hline
		$\mu^+\to e^+\gamma$ & BR $\leq 4.3\times 10^{-13}$ \cite{TheMEG:2016wtm} & $6\times 10^{-14}$~\cite{Baldini:2018nnn} &$|\Delta (L_e-L_\mu)|=2$\\
	\hline
	$\tau^-\to \mu^-\gamma$ & BR $\leq 4.4\times 10^{-8}$ \cite{Aubert:2009ag} & $10^{-9}$~\cite{Kou:2018nap} &$|\Delta (L_\tau-L_\mu)|=2$\\
	\hline
	$\tau^-\to e^-\gamma$ & BR $\leq 3.3\times 10^{-8}$ \cite{Aubert:2009ag} & $3\times 10^{-9}$~\cite{Kou:2018nap} &$|\Delta (L_\tau-L_e)|=2$\\
	\hline
	$\mu^+ \to e^+ e^- e^+$ & BR $\leq 1.0\times 10^{-12}$  \cite{Bellgardt:1987du} & $10^{-16}$~\cite{Blondel:2013ia} & $|\Delta (L_e-L_\mu)|=2$\\
	\hline
	$\tau^-\to e^- e^+ e^-$ & BR $\leq 2.7\times 10^{-8}$ \cite{Hayasaka:2010np} & $4\times 10^{-10}$~\cite{Kou:2018nap} &$|\Delta (L_\tau-L_e)|=2$\\
	\hline
	$\tau^-\to \mu^- \mu^+ \mu^-$ & BR $\leq 2.1\times 10^{-8}$ \cite{Hayasaka:2010np} & $4\times10^{-10}$~\cite{Kou:2018nap} & $|\Delta (L_\tau-L_\mu)|=2$\\
	\hline
	$\tau^-\to e^+ \mu^- \mu^-$ & BR $\leq 1.7\times 10^{-8}$ \cite{Hayasaka:2010np} & $3\times10^{-10}$~\cite{Kou:2018nap} &$|\Delta (L_\tau+L_e-2L_\mu)|=6$\\
	\hline
	$\tau^-\to \mu^+ e^- e^-$ & BR $\leq1.5\times 10^{-8}$ \cite{Hayasaka:2010np} & $3\times 10^{-10}$~\cite{Kou:2018nap} &$|\Delta (L_\tau+L_\mu-2L_e)|=6$\\
	\hline
	$\tau^-\to \mu^- e^+ e^-$ & BR $\leq 1.8\times 10^{-8}$ \cite{Hayasaka:2010np} & $3\times10^{-10}$~\cite{Kou:2018nap} &$|\Delta (L_\tau-L_\mu)|=2$\\
	\hline
	$\tau^-\to e^- \mu^+ \mu^-$ & BR $\leq 2.7\times 10^{-8}$ \cite{Hayasaka:2010np}& $4\times 10^{-10}$~\cite{Kou:2018nap} & $|\Delta (L_\tau-L_e)|=2$ \\
	\hline
\end{tabular}

\vspace{0.5cm}

\begin{tabular}{|c|c|c|}\hline
	process &experimental constraint & lepton flavor symmetry \\\hline
	$\mu^+e^- \to \mu^- e^+$ & $P_{M\bar M}(0.1 T) \leq 8.3\times 10^{-11}$ \cite{Willmann:1998gd} &   $|\Delta (L_\mu-L_e)|=4$\\\hline
	\end{tabular}

\vspace{0.5cm}

\begin{tabular}{|c|c|}\hline
	measurement & lepton flavor symmetry \\\hline
	$a_e=(1159652180.91 \pm0.26) \times 10^{-12}$  \cite{Mohr:2015ccw} & $|\Delta L_e|=0$\\\hline
	$a_\mu=(11659208.9 \pm5.4 \pm3.3)\times 10^{-10}$  \cite{Bennett:2006fi}  & $|\Delta L_\mu| = 0$\\
		\hline
	\end{tabular}
\caption{Relevant constraints from low-energy precision experiments testing LFV. The top table lists constraints from decay, the center one muonium-antimuonium conversion, and the bottom one shows the current experimental values for the anomalous magnetic moments of the different leptons. The projected sensitivity of the Belle II experiment to LFV $\tau$ decays has been read off from Fig.~177 of Ref.~\cite{Kou:2018nap}.}
	\label{tab:LE}
\end{table}	
The trilepton decay constraints are obtained by a straightforward tree-level calculation and are given as a ratio of the three-body decay branching ratio from new physics contribution and the dominant leptonic SM branching ratio. We neglect loop-level contributions to trilepton decays, because they are generally suppressed. Box diagrams become relevant and similar in size to the tree-level contribution, if one of the couplings is close to its perturbative limit $\sim4\pi$. Although this leads to an $O(1)$ correction to the branching ratio, the tree-level calculation still provides a good estimate and is sufficiently precise for the following discussion. In some cases additional trilepton decay modes are induced via box diagrams. For example, with non-zero couplings $y_1^{e\mu}y_1^{e\tau}$, the $\tau$ decay to $3\mu$ is induced in addition to the tree-level decay $\tau^-\to \mu^-e^+e^-$. Given the comparable experimental sensitivity of leptonic LFV $\tau$ decays, the loop-induced decays do not provide stronger constraints.
For the radiative LFV decays and the anomalous magnetic moments we use the general formulas provided by Lavoura in Ref.~\cite{Lavoura:2003xp}.
The experimental measurement of the muon anomalous magnetic moment differs from the SM prediction by more than $3\sigma$~\cite{Tanabashi:2018oca} and the electron magnetic moment shows deviations at $2.5\sigma$~\cite{Parker:2018vye}.
Hence we demand the muon anomalous magnetic moment to be only within $4\sigma$ of the experimental measurement, while we require the anomalous magnetic moment of the electron to be within $3\sigma$ of the experimental measurement.

The muonium-antimuonium conversion probability in the absence of magnetic fields is given by $P_{M\bar M} = |\delta|^2/2\Gamma_\mu^2$~\cite{Feinberg:1961zz,Feinberg:1961zza} with the transition matrix element $\langle\bar M| H| M\rangle=\delta/2$ and the muon decay rate $\Gamma_\mu$.
For effective left-handed (right-handed) vector interactions such as $H=C [\bar \mu \gamma_\rho P_{L,R} e] [\bar \mu \gamma^\rho P_{L,R} e]$, we find $\delta=4 C /\pi a^3$ \cite{Feinberg:1961zz} irrespective whether the muonium and antimuonium are both in a triplet or singlet state. The Bohr radius of muonium is $a=(\alpha_{\rm em} m_e m_\mu/(m_e+m_\mu))^{-1}$. The effective Hamiltonian $H=D [\bar \mu \gamma_\rho P_{L,R} e] [\bar \mu \gamma^\rho P_{R,L} e]$ leads to $\delta=-2 D/\pi a^3$ for the triplet state and $\delta=6 D/\pi a^3$ for the singlet state~\cite{Fujii:1992np}.
There is no contribution from tensor interactions and the relevant scalar and pseudoscalar interactions can be recast in terms of the vector interactions using Fierz rearrangements.
Typically there are magnetic fields in the experimental setup which suppress the conversion probability because the degeneracy of the energy levels in $M$ and $\bar M$ is lifted. Thus the conversion probability in presence of a magnetic field is $P_{M\bar M}(B) = S_B P_{M\bar M}$, where $S_B$ denotes a suppression factor. The suppression factors for a magnetic field of $0.1$ Tesla are
$S_B=0.35$ for same $[LL,RR]$ chiralities and $S_B=0.78$ for opposite $[RL]$ chiralities.\cite{Horikawa:1995ae,Willmann:1998gd}

\begin{table}[tbp!]
\begin{center}
\begin{tabular}{|c|c|c|}
\hline
process  & $H_2^0$ & $H_1^{(\prime)0}$
\\ \hline
$\mu^+\to e^+\gamma$ & $|(y_2^\dagger y_2)^{e\mu}|<1.6\times 10^{-9} m_{h_2}^2$ & $|(y_1^{(\prime)\dagger} y_1^{(\prime)})^{e\mu}|<5.5\times 10^{-10} m_{H_1^{(\prime)}}^2$
\\
& [$5.8\times 10^{-10} m_{h_2}^2$] & [$2.1\times 10^{-10} m_{H_1^{(\prime)}}^2$]
\\ \hline
$\tau^-\to \mu^-\gamma$ & $|(y_2^\dagger y_2)^{\mu\tau}|<1.2\times 10^{-6} m_{h_2}^2$ & $|(y_1^{(\prime)\dagger} y_1^{(\prime)})^{\mu\tau}|<4.2\times 10^{-7} m_{H_1^{(\prime)}}^2$
\\
& [$1.8\times 10^{-7} m_{h_2}^2$] & [$6.4\times 10^{-8} m_{H_1^{(\prime)}}^2$]
\\ \hline
$\tau^-\to e^-\gamma$ & $|(y_2^\dagger y_2)^{e\tau}|<1.0\times 10^{-6} m_{h_2}^2$ & $|(y_1^{(\prime)\dagger} y_1^{(\prime)})^{e\tau}|<3.6\times 10^{-7} m_{H_1^{(\prime)}}^2$
\\
& [$3.1\times 10^{-7} m_{h_2}^2$] & [$1.1\times 10^{-7} m_{H_1^{(\prime)}}^2$]
\\ \hline
$\mu^+\to e^+e^-e^+$ & $|y_2^{ee} y_2^{e\mu}|<4.7\times 10^{-11} m_{h_2}^2$ & $|y_1^{(\prime)ee} y_1^{(\prime)e\mu}|<1.9\times 10^{-11} m_{H_1^{(\prime)}}^2$
\\
& [$4.7\times 10^{-13} m_{h_2}^2$] & [$1.9\times 10^{-13} m_{H_1^{(\prime)}}^2$]
\\ \hline
$\tau^-\to e^-e^+e^-$ & $|y_2^{ee} y_2^{e\tau}|<1.8\times 10^{-8} m_{h_2}^2$ & $|y_1^{(\prime)ee} y_1^{(\prime)e\tau}|<7.4\times 10^{-9} m_{H_1^{(\prime)}}^2$
\\
& [$2.2\times 10^{-9} m_{h_2}^2$] & [$9.0\times 10^{-10} m_{H_1^{(\prime)}}^2$]
\\ \hline
$\tau^-\to \mu^-\mu^+\mu^-$ & $|y_2^{\mu\mu} y_2^{\mu\tau}|<1.6\times 10^{-8} m_{h_2}^2$ & $|y_1^{(\prime)\mu\mu} y_1^{(\prime)\mu\tau}|<6.6\times 10^{-9} m_{H_1^{(\prime)}}^2$
\\
& [$2.2\times 10^{-9} m_{h_2}^2$] & [$9.1\times 10^{-10} m_{H_1^{(\prime)}}^2$]
\\ \hline
$\tau^-\to e^+\mu^-\mu^-$ & $|y_2^{e\mu}y_2^{\mu\tau}|<1.5\times 10^{-8} m_{h_2}^2$  & $|y_1^{(\prime)e\mu} y_1^{(\prime)\mu\tau}|<6.0\times 10^{-9} m_{H_1^{(\prime)}}^2$
\\
& [$1.9\times 10^{-9} m_{h_2}^2$] & [$7.9\times 10^{-10} m_{H_1^{(\prime)}}^2$]
\\ \hline
$\tau^-\to \mu^+e^-e^-$ & $|y_2^{e\mu}y_2^{e\tau}|<1.4\times 10^{-8} m_{h_2}^2$  & $|y_1^{(\prime)e\mu} y_1^{(\prime)e\tau}|<5.6\times 10^{-9} m_{H_1^{(\prime)}}^2$
\\
& [$1.9\times 10^{-9} m_{h_2}^2$] & [$7.9\times 10^{-10} m_{H_1^{(\prime)}}^2$]
\\ \hline
$\tau^-\to \mu^-e^+e^-$ & $(|y_2^{ee}y_2^{\mu\tau}|^2+|y_2^{e\mu}y_2^{e\tau}|^2)^{1/2}$  & $(|y_1^{(\prime)ee}y_1^{(\prime)\mu\tau}|^2+|y_1^{(\prime)e\mu}y_1^{(\prime)e\tau}|^2+$
\\
& $<1.5\times 10^{-8} m_{h_2}^2$ & $4y_1^{(\prime)ee}y_1^{(\prime)\mu\tau}y_1^{(\prime)e\mu}y_1^{(\prime)e\tau})^{1/2}<1.1\times 10^{-8} m_{H_1^{(\prime)}}^2$
\\
& [$1.9\times 10^{-9} m_{h_2}^2$] & [$1.4\times 10^{-9} m_{H_1^{(\prime)}}^2$]
\\ \hline
$\tau^-\to e^-\mu^+\mu^-$ & $(|y_2^{\mu\mu}y_2^{e\tau}|^2+|y_2^{e\mu}y_2^{\mu\tau}|^2)^{1/2}$ & $(|y_1^{(\prime)\mu\mu}y_1^{(\prime)e\tau}|^2+|y_1^{(\prime)e\mu}y_1^{(\prime)\mu\tau}|^2+$
\\
& $<1.8\times 10^{-8} m_{h_2}^2$ & $4y_1^{(\prime)\mu\mu}y_1^{(\prime)e\tau}y_1^{(\prime)e\mu}y_1^{(\prime)\mu\tau})^{1/2}<1.3 \times 10^{-8}\ m_{H_1^{(\prime)}}^2$
\\
& [$2.2\times 10^{-9} m_{h_2}^2$] & [$1.6\times 10^{-9} m_{H_1^{(\prime)}}^2$]
\\ \hline
$\mu^+ e^-\to \mu^- e^+$ & $|y_2^{e\mu}|^2 < 1.6\times 10^{-7}\ m_{h_2}^2$ & $|y_1^{(\prime)e\mu}|^2 < 2.0\times 10^{-7}\ m_{H_1^{(\prime)}}^2$
\\ \hline
$a_e\; [3\sigma]$ & $|(y_2^\dagger y_2)^{ee}| < 1.6\times 10^{-3}\ m_{h_2}^2$ & $|(y_1^{(\prime)\dagger} y_1^{(\prime)})^{ee}| < 1.5\times 10^{-4}\ m_{H_1^{(\prime)}}^2$ \\\hline
$a_\mu\; [4\sigma]$ & $|(y_2^\dagger y_2)^{\mu\mu}| < 8.4\times 10^{-6}\ m_{h_2}^2$ & $|(y_1^{(\prime)\dagger} y_1^{(\prime)})^{\mu\mu}| < 6.1\times 10^{-5}\ m_{H_1^{(\prime)}}^2$ \\ \hline
\end{tabular}
\caption{Constrained couplings in units of ${\rm GeV}^{-2}$ for $H_2^0$ and $H_1^{(\prime)0}$. Here we assume all the CLFV couplings are real and symmetric, and $m_{h_2}=m_{a_2}$ for $H_2^0$. Future sensitivities	are indicated inside square brackets.
}
\label{tab:couplingsH}
\end{center}
\end{table}
LFV $Z$-boson decays do not lead to competitive constraints: They are several orders of magnitude weaker, because they are induced at loop-level and the current experimental limits are of order $10^{-5}$ for $Z\to \tau \ell$. Although the sensitivity will be improved by 3-6 orders of magnitude at the CEPC~\cite{CEPCStudyGroup:2018ghi}, the future sensitivity of Belle II to trilepton $\tau$ decays will be more sensitive for our parameters of interest.

From the experimental constraints in Table~\ref{tab:LE} we obtain constraints on the different combinations of CLFV couplings for $\Delta L=0$ and $\Delta L=2$ interactions which are reported in Tables~\ref{tab:couplingsH} and \ref{tab:couplingsD}, respectively. Future sensitivities by Belle II~\cite{Kou:2018nap}, the MEG upgrade~\cite{Baldini:2013ke}, and Mu3E~\cite{Blondel:2013ia} are indicated inside square brackets.
The detailed formulas for low-energy precision constraints are collected in the Appendix.
Note that, for the numerical results in this and the next sections, we assume all the CLFV couplings to be real and symmetric. For $H_2^0$ we additionally assume equal masses for the scalar $h_2$ and the pseudoscalar $a_2$, $m_{h_2}=m_{a_2}$.

\begin{table}[tbp!]
\begin{center}
\begin{tabular}{|c|c|c|}
\hline
process  & $\Delta_{1,3}^{++}$  & $\Delta_2^{++}$
\\ \hline
$\mu^+\to e^+\gamma$ & $|(\lambda_{1,3}^\dagger \lambda_{1,3})^{e\mu}|<1.1\times 10^{-9}m_{\Delta_{1,3}}^2$ & $|(\lambda_2^\dagger \lambda_2)^{e\mu}|<1.6\times 10^{-10}m_{\Delta_2}^2$
\\
   & [$4.1\times 10^{-10}m_{\Delta_{1,3}}^2$] & [$5.9\times 10^{-11}m_{\Delta_2}^2$]
\\ \hline
$\tau^-\to \mu^-\gamma$ & $|(\lambda_{1,3}^\dagger \lambda_{1,3})^{\mu\tau}|<8.4\times 10^{-7}m_{\Delta_{1,3}}^2$ & $|(\lambda_2^\dagger \lambda_2)^{\mu\tau}|<1.2\times 10^{-7}m_{\Delta_2}^2$
\\
   & [$1.3\times 10^{-7}m_{\Delta_{1,3}}^2$] & [$1.8\times 10^{-8}m_{\Delta_2}^2$]
\\ \hline
$\tau^-\to e^-\gamma$ & $|(\lambda_{1,3}^\dagger \lambda_{1,3})^{e\tau}|<7.2\times 10^{-7}m_{\Delta_{1,3}}^2$ &  $|(\lambda_2^\dagger \lambda_2)^{e\tau}|<1.0\times 10^{-7}m_{\Delta_2}^2$
\\
   & [$2.2\times 10^{-7}m_{\Delta_{1,3}}^2$] & [$3.1\times 10^{-8}m_{\Delta_2}^2$]
\\ \hline
$\mu^+\to e^+e^-e^+$ & $|\lambda_{1,3}^{ee} \lambda_{1,3}^{e\mu}|<2.3\times 10^{-11} m_{\Delta_{1,3}}^2$ & $|\lambda_2^{ee} \lambda_2^{e\mu}|<2.3\times 10^{-11} \ m_{\Delta_2}^2$
\\
   & [$2.3\times 10^{-13}m_{\Delta_{1,3}}^2$] & [$2.3\times 10^{-13}m_{\Delta_2}^2$]
\\ \hline
$\tau^-\to e^-e^+e^-$ & $|\lambda_{1,3}^{ee} \lambda_{1,3}^{e\tau}|<9.1\times 10^{-9} m_{\Delta_{1,3}}^2$ & $|\lambda_2^{ee} \lambda_2^{e\tau}|<9.1\times 10^{-9} \ m_{\Delta_2}^2$
\\
   & [$1.1\times 10^{-9} m_{\Delta_{1,3}}^2$] & [$1.1\times 10^{-9}m_{\Delta_2}^2$]
\\ \hline
$\tau^-\to \mu^-\mu^+\mu^-$ & $|\lambda_{1,3}^{\mu\mu} \lambda_{1,3}^{\mu\tau}|<8.1\times 10^{-9} m_{\Delta_{1,3}}^2$ & $|\lambda_2^{\mu\mu} \lambda_2^{\mu\tau}|<8.1\times 10^{-9} \ m_{\Delta_2}^2$
\\
   & [$1.1\times 10^{-9} m_{\Delta_{1,3}}^2$] & [$1.1\times 10^{-9}m_{\Delta_2}^2$]
\\ \hline
$\tau^-\to e^+\mu^-\mu^-$ & $|\lambda_{1,3}^{e\tau} \lambda_{1,3}^{\mu\mu}|<7.3\times 10^{-9} m_{\Delta_{1,3}}^2$ & $|\lambda_2^{e\tau} \lambda_2^{\mu\mu}|<7.3\times 10^{-9} \ m_{\Delta_2}^2$
\\
   & [$9.7\times 10^{-10} m_{\Delta_{1,3}}^2$] & [$9.7\times 10^{-10}m_{\Delta_2}^2$]
\\ \hline
$\tau^-\to \mu^+e^-e^-$ & $|\lambda_{1,3}^{ee} \lambda_{1,3}^{\mu\tau}|<6.8\times 10^{-9} m_{\Delta_{1,3}}^2$ & $|\lambda_2^{ee} \lambda_2^{\mu\tau}|<6.8\times 10^{-9} \ m_{\Delta_2}^2$
\\
   & [$9.7\times 10^{-10} m_{\Delta_{1,3}}^2$] & [$9.7\times 10^{-10}m_{\Delta_2}^2$]
\\ \hline
$\tau^-\to \mu^-e^+e^-$ & $|\lambda_{1,3}^{e\mu} \lambda_{1,3}^{e\tau}|<5.3\times 10^{-9} m_{\Delta_{1,3}}^2$ & $|\lambda_2^{e\mu} \lambda_2^{e\tau}|<5.3\times 10^{-9} \ m_{\Delta_2}^2$
\\
   & [$6.8\times 10^{-10} m_{\Delta_{1,3}}^2$] & [$6.8\times 10^{-10}m_{\Delta_2}^2$]
\\ \hline
$\tau^-\to e^-\mu^+\mu^-$ & $|\lambda_{1,3}^{e\mu} \lambda_{1,3}^{\mu\tau}|<6.5\times 10^{-9} m_{\Delta_{1,3}}^2$ & $|\lambda_2^{e\mu} \lambda_2^{\mu\tau}|<6.5\times 10^{-9} \ m_{\Delta_2}^2$
\\
   & [$7.9\times 10^{-10} m_{\Delta_{1,3}}^2$] & [$7.9\times 10^{-10}m_{\Delta_2}^2$]
\\ \hline
$\mu^+ e^-\to \mu^- e^+$ & $|\lambda_{1,3}^{ee}\lambda_{1,3}^{\mu\mu}| < 2.0\times 10^{-7}\ m_{\Delta_{1,3}}^2$ &$|\lambda_2^{ee}\lambda_2^{\mu\mu}| < 7.8\times 10^{-8}\ m_{\Delta_2}^2$
\\ \hline
$a_e\;[3\sigma]$ & $|(\lambda_{1,3}^\dagger \lambda_{1,3})^{ee}| < 3.1\times 10^{-4}\ m_{\Delta_{1,3}}^2$ &$|(\lambda_2^\dagger \lambda_2)^{ee}| < 2.3\times 10^{-4}\ m_{\Delta_2}^2$\\ \hline
$a_\mu\; [4\sigma]$ & $|(\lambda_{1,3}^\dagger \lambda_{1,3})^{\mu\mu}| < 1.2\times 10^{-4}\ m_{\Delta_{1,3}}^2$ &$|(\lambda_2^\dagger \lambda_2)^{\mu\mu}| < 1.2\times 10^{-6}\ m_{\Delta_2}^2$\\ \hline
\end{tabular}
\end{center}
\caption{Constrained couplings in units of ${\rm GeV}^{-2}$ for $\Delta_1^{++}$ ($\Delta_3^{++}$) and $\Delta_2^{++}$. Here we assume all the CLFV couplings are real and symmetric.
 Future sensitivities are indicated inside square brackets.
}
\label{tab:couplingsD}
\end{table}

\subsection{Current collider constraints}
\label{sec:collcons}
There are already constraints from experiments at the LEP and the LHC colliders. In Ref.~\cite{Abdallah:2005ph} the DELPHI collaboration interpreted their searches for $e^+e^-\to \ell^+\ell^-$ in terms of leptonic dimension 6 operators defined by the effective Lagrangian
\begin{equation}
\mathcal{L}_{eff} = \frac{g^2}{(1+\delta_{e\ell})\Lambda^2} \sum_{i,j=L,R} \eta_{ij} \bar e_i\gamma_\mu e_i\bar \ell_j \gamma^\mu \ell_j\;,
\end{equation}
where $\Lambda$ denotes the scale of the effective operator, $g$ is the coupling and $\eta_{ij}$ parameterizes which operators are considered at a given time and the relative sign of the operators in order to distinguish constructive (destructive) interference with the SM contribution. The analysis sets the coupling to $g^2=4\pi$ to obtain conservative limits on the new physics scale, which are summarized in Tab.~30 of Ref.~\cite{Abdallah:2005ph}. We list the translated limits for masses well above the center-of-mass energy of LEP, $\sqrt{s}\sim 130-207$ GeV, in Tab.~\ref{tab:LEPbounds}. The analysis of contact interactions in Ref.~\cite{Abdallah:2005ph} does not directly apply to $\Delta_{2\mu}^{++}$, because the induced effective interactions do not fall into the class of effective interactions considered in Ref.~\cite{Abdallah:2005ph}. Note that these limits are only valid when the new particle mass is much greater than $\sqrt{s}$.
To make it valid for any masses, we should replace the mass in Tab.~\ref{tab:LEPbounds} by $(s\cos\theta/2+m^2)^{-1/2}$ after averaging over the scattering angle $\langle\cos\theta\rangle\simeq 1/2$.
Nevertheless, the LEP limits are less restrictive than the low-energy precision constraints.

\begin{table}[htb]\centering
\begin{tabular}{|c|c|c|c|}\hline
 & $e^+e^- \to e^+ e^-$ & $e^+e^- \to \mu^+\mu^-$ & $e^+e^- \to \tau^+\tau^-$\\\hline
$H_1$ & $|y_1^{ee}|\leq 6.7\times 10^{-4} m_{H_1}$ & $|y_1^{e\mu}|\leq 4.9\times 10^{-4} m_{H_1}$ & $|y_1^{e\tau}|\leq 4.5\times 10^{-4} m_{H_1}$
\\\hline
$H_1^\prime$ & $|y_1^{\prime ee}|\leq 6.8\times 10^{-4} m_{H_1^\prime}$ & $|y_1^{\prime e\mu}|\leq 5.1\times 10^{-4} m_{H_1^\prime}$& $|y_1^{\prime e\tau}|\leq 4.7\times 10^{-4} m_{H_1^\prime}$
\\\hline
$H_2$ & $|y_2^{ee}|\leq 5.3\times 10^{-4} m_{h_2}$ & $|y_2^{e\mu}|\leq 2.5\times 10^{-3} m_{h_2}$& $|y_2^{e\tau}|\leq 2.4\times 10^{-3} m_{h_2}$
\\\hline
$\Delta_1^{++}$ & $|\lambda_1^{ee}|\leq 6.8\times 10^{-4} m_{\Delta_1}$ & $|\lambda_1^{e\mu}|\leq 3.6\times 10^{-4} m_{\Delta_1}$ & $|\lambda_1^{e\tau}|\leq 3.3\times 10^{-4} m_{\Delta_1}$
\\\hline
$\Delta_3^{++}$ & $|\lambda_3^{ee}|\leq 6.7\times 10^{-4} m_{\Delta_3}$ & $|\lambda_3^{e\mu}|\leq 3.4\times 10^{-4} m_{\Delta_3}$ & $|\lambda_3^{e\tau}|\leq 3.2\times 10^{-4} m_{\Delta_3}$
\\\hline
\end{tabular}
\caption{LEP limits for masses well above the center of mass energy $\sqrt{s}\sim 130-207$ GeV. }
\label{tab:LEPbounds}
\end{table}

The ATLAS and CMS experiments
searched for electroweak pair production of doubly-charged scalars with subsequent decay to $e^\pm e^\pm$, $\mu^\pm \mu^\pm$, $e^\pm \mu^\pm$ pairs. Currently the most stringent limits by the ATLAS~\cite{ATLAS:2017iqw} are $m_{\Delta^{++}_{1(3)}}\geq 320 (450)$ GeV assuming BR$(\Delta^{++}_{1,3} \to \ell^+ \ell^+)\geq 10\%$. The limits from searches for $\tau$ lepton final state are less stringent by about 200 GeV, compared to the above constraints~\cite{CMS:2017pet}.

\subsection{Other constraints}
Although we focus on the models defined in Eqs.~\eqref{eq:delL0} and \eqref{eq:delL2} we briefly comment on constraints from embedding the new bosons in complete SM multiplets.

There is an additional left-handed vector interaction of $H_1$ with neutrinos. This introduces new contributions to leptonic lepton decays $\ell \to \ell^\prime \nu \bar \nu^\prime$, in particular to leptonic muon decay which is used to extract the Fermi constant. Given the stringent constraints from CLFV in Tab.~\ref{tab:couplingsH} the modifications to $\ell\to\ell^\prime \nu \bar\nu^\prime$ are small compared to the SM contribution and thus we neglect them in the processes which we are considering. Similarly neutrino trident production~\cite{Altmannshofer:2014pba}, the production of a $\mu^+\mu^-$ pair from the a neutrino scattering off the Coulomb field of a nucleus, provides a constraint. An order of magnitude estimate of the constraint on the $H_1$ gauge coupling, $y_1\lesssim 0.01$ for $m_{H_1}\simeq 10$ GeV, shows that the constraints from low-energy precision experiments outperform the one from neutrino trident production for the parameter region of interest. The same conclusion holds for the other particles which induce processes with neutrinos.

Several of the other particles, $H_2$, $\Delta_3^{++}$ and $\Delta_{2\mu}^{++}$ are accompanied by a singly-charged scalar which are searched for at colliders. In particular the singly-charged scalar of a second electroweak doublet has been searched for at the LHC under the assumption that it couples to quarks and thus these searches do not apply in this case.

The additional particles in the multiplets also contribute to the radiative LFV decay and may interfere constructively like for $\Delta_3$ or destructively like for $H_2$. As the LFV tree-level trilepton decays are more stringent and do not receive any additional contributions, we will only show the limits from LFV tree-level trilepton decays in the following.

\section{Sensitivity of future lepton colliders to the CLFV}
\label{sec:sen}

The CLFV processes can happen through the scattering of either opposite-sign~\cite{Dev:2017ftk,Sui:2017qra,Dev:2018upe} or same-sign~\cite{Cuypers:1997qg,Rodejohann:2010bv} leptons, i.e. $\ell_0^+\ell_0^-\to \ell^\pm_i\ell^\mp_j$ or $\ell_0^-\ell_0^-\to \ell^-_i\ell^-_j$ with $\ell_0=e,\mu$ and $\ell_i,\ell_j=e,\mu,\tau$,
mediated by the new bosonic particles in Eqs.~(\ref{eq:delL0}) and (\ref{eq:delL2}) at tree level. The mediators are off-shell in both s and t channels, except for the resonance case where the mediator mass is close to the c.m.~energy in s channel. In this section, we analyze the sensitivity of proposed lepton colliders to the CLFV couplings and directly compare with the constrained couplings from low-energy precision observables. The following results are mainly for $e^+e^-$ or $e^-e^-$ colliders. Actually, the explored lepton flavors and the dedicated CLFV processes at a muon collider can be easily obtained by swapping $e$ and $\mu$ flavors in the couplings and both initial and final states of processes at $e^+e^-$ or $e^-e^-$ colliders.

We apply basic cuts $p_T>10$ GeV and $|\eta|<2.5$ on the leptons in final state
and assume 10 discovered signal events. The total width of the mediating
particle is set to be 10 GeV~\cite{Dev:2017ftk}. If we take tau efficiency of
$60\%$~\cite{Baer:2013cma}, the sensitivity to the combination of Yukawa couplings in the scattering amplitude will become weaker by
$77\%$ and $60\%$ for final states with one and two tau leptons, respectively.
Note that we do not intend to discriminate the chiral nature of the couplings of the mediating particles, i.e.~we only consider the total number of signal events. Thus, the following
results for $H_1^0$ and $\Delta_3^{++}$ which only couple to left-handed leptons are
the same as those for $H_1^{\prime 0}$ and $\Delta_1^{++}$ with only
couplings to right-handed leptons, respectively. Different chirality interactions can be
distinguished by measuring the angular distribution with polarized
beams~\cite{Han:2013mra,Nomura:2017abh,JohnNg}.

\subsection{Opposite-sign lepton collision}

Table~\ref{tab:opposite-sign} summarizes the induced CLFV channels at an $e^+e^-$ collider and the relevant couplings in either the $\Delta L=0$ or $\Delta L=2$ Lagrangian.
One can see that, at an opposite-sign lepton collider, the $\Delta L=0$ interactions can mediate $e^+e^-\to e^\pm \mu^\mp, e^\pm \tau^\mp$ processes in both s and t channel. The $e^+e^-\to \mu^\pm \tau^\mp$ process without electron or positron in final states can happen in either s or t channel, governed by different coupling configurations. The $\Delta L=2$ interactions only occur in t channel, as a result of the fermion flow of charge.

\begin{table}[htbp!]
\begin{center}
\begin{tabular}{|c|c|c|c|}
        \hline
        CLFV channel & flavor $ij,i'j'$ & $\Delta L=0$ & $\Delta L=2$\\
        \hline
        $e^+e^-\to e^\pm\mu^\mp$ & $ee,e\mu$ & s+t & t \\
        \hline
        $e^+e^-\to e^\pm\tau^\mp$ & $ee,e\tau$ & s+t & t \\
        \hline
        $e^+e^-\to \mu^\pm\tau^\mp$ & $ee,\mu\tau$ & s & - \\
        \hline
        $e^+e^-\to \mu^\pm\tau^\mp$ & $e\mu,e\tau$ & t & t \\
        \hline
\end{tabular}
\end{center}
\caption{CLFV channels for probing coupling $y^{ij}y^{i'j'}$ or $\lambda^{ij}\lambda^{i'j'}$ via $\Delta L=0$ or $\Delta L=2$ interaction at $e^+e^-$ collider, while $e\leftrightarrow\mu$ for $\mu^+\mu^-$ collider.}
\label{tab:opposite-sign}
\end{table}

\begin{figure}[htbp!]\centering
\includegraphics[scale=1,width=6.5cm]{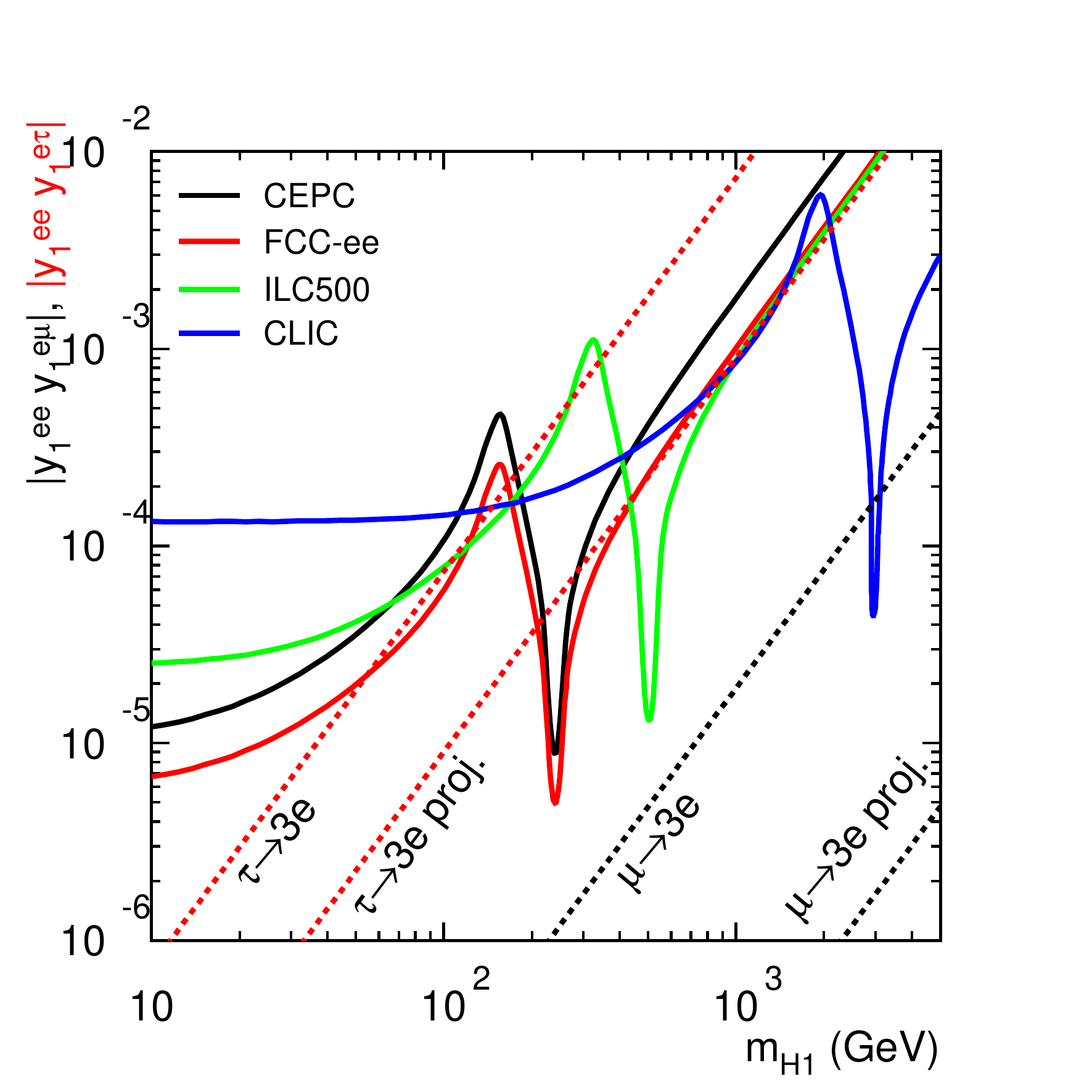}
\includegraphics[scale=1,width=6.5cm]{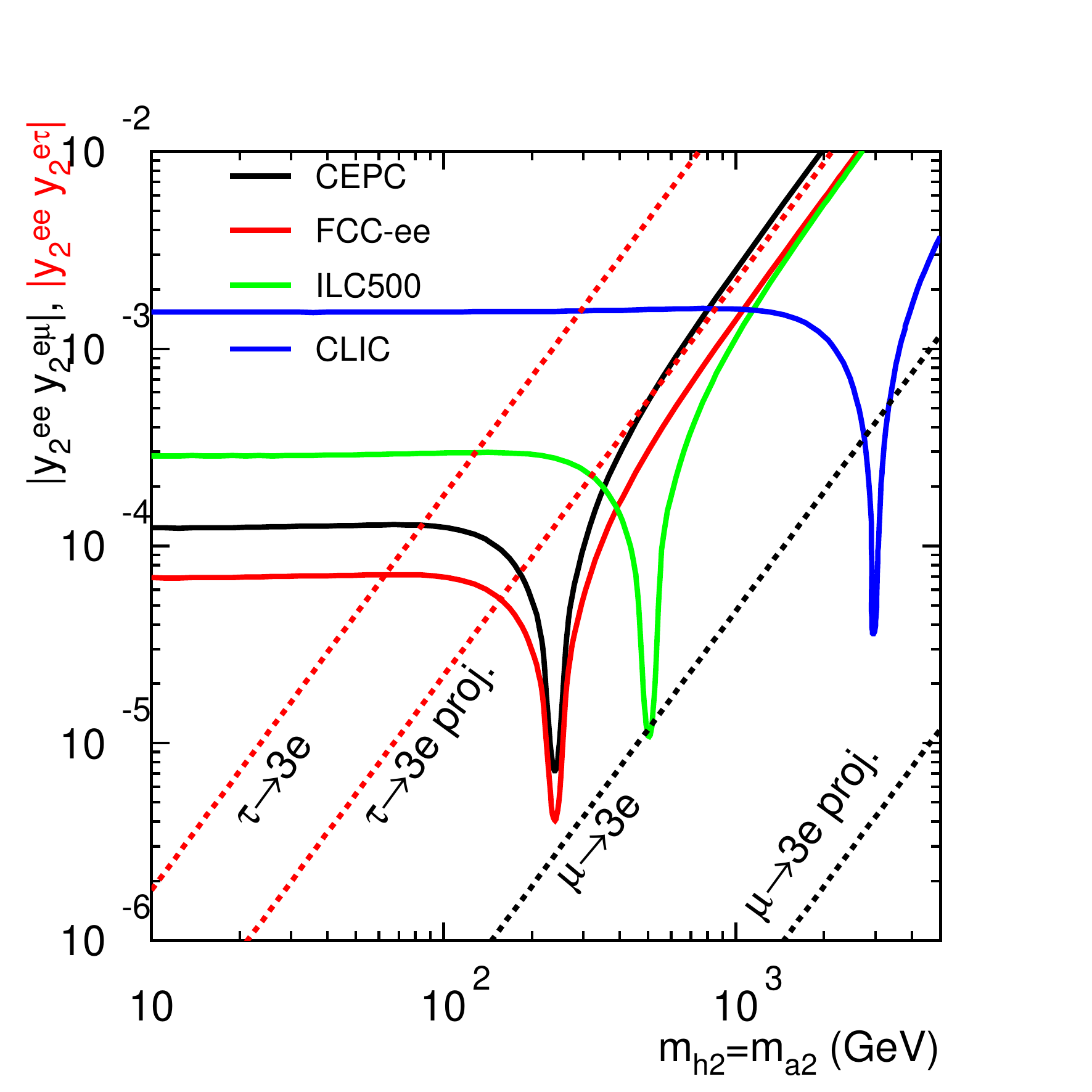}
\includegraphics[scale=1,width=6.5cm]{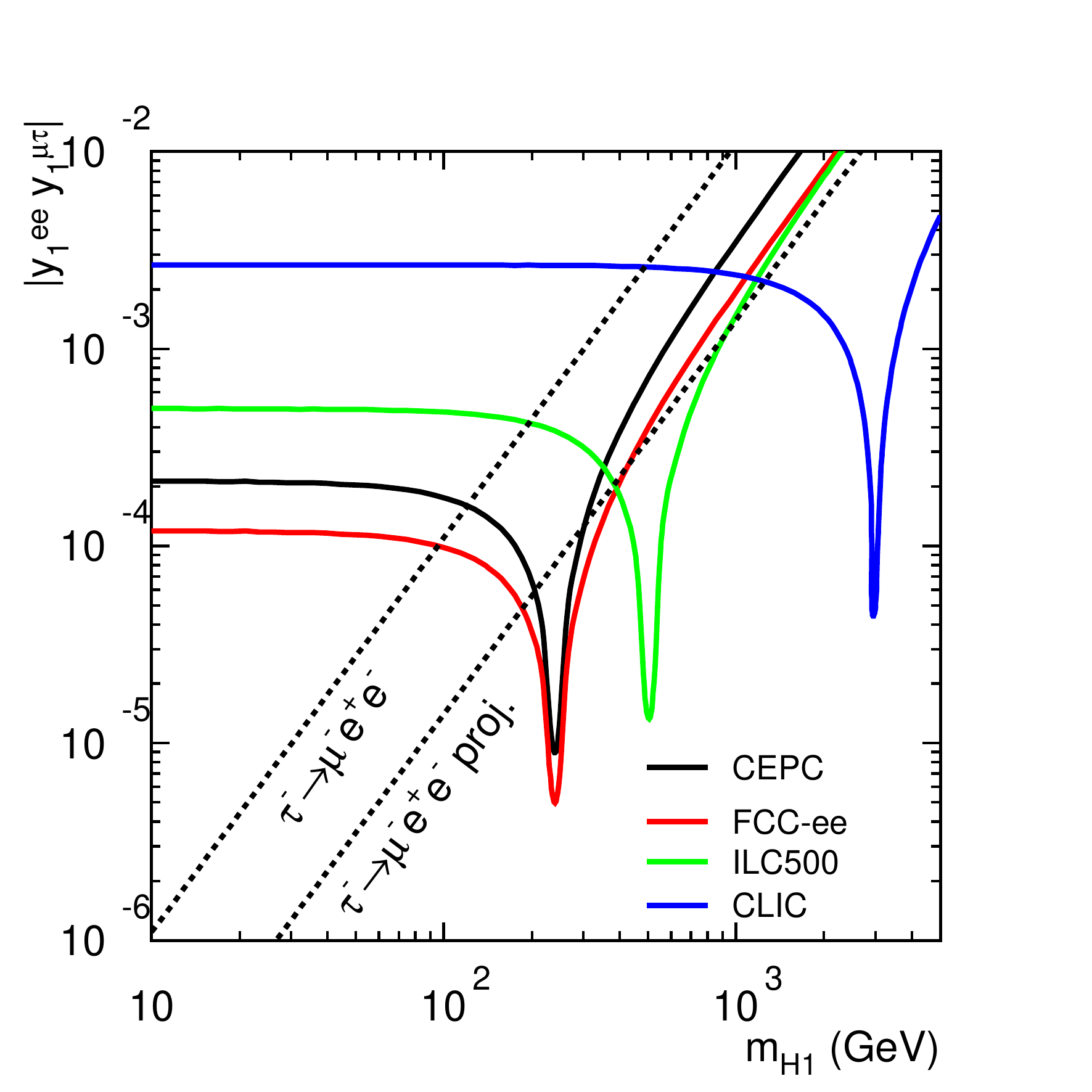}
\includegraphics[scale=1,width=6.5cm]{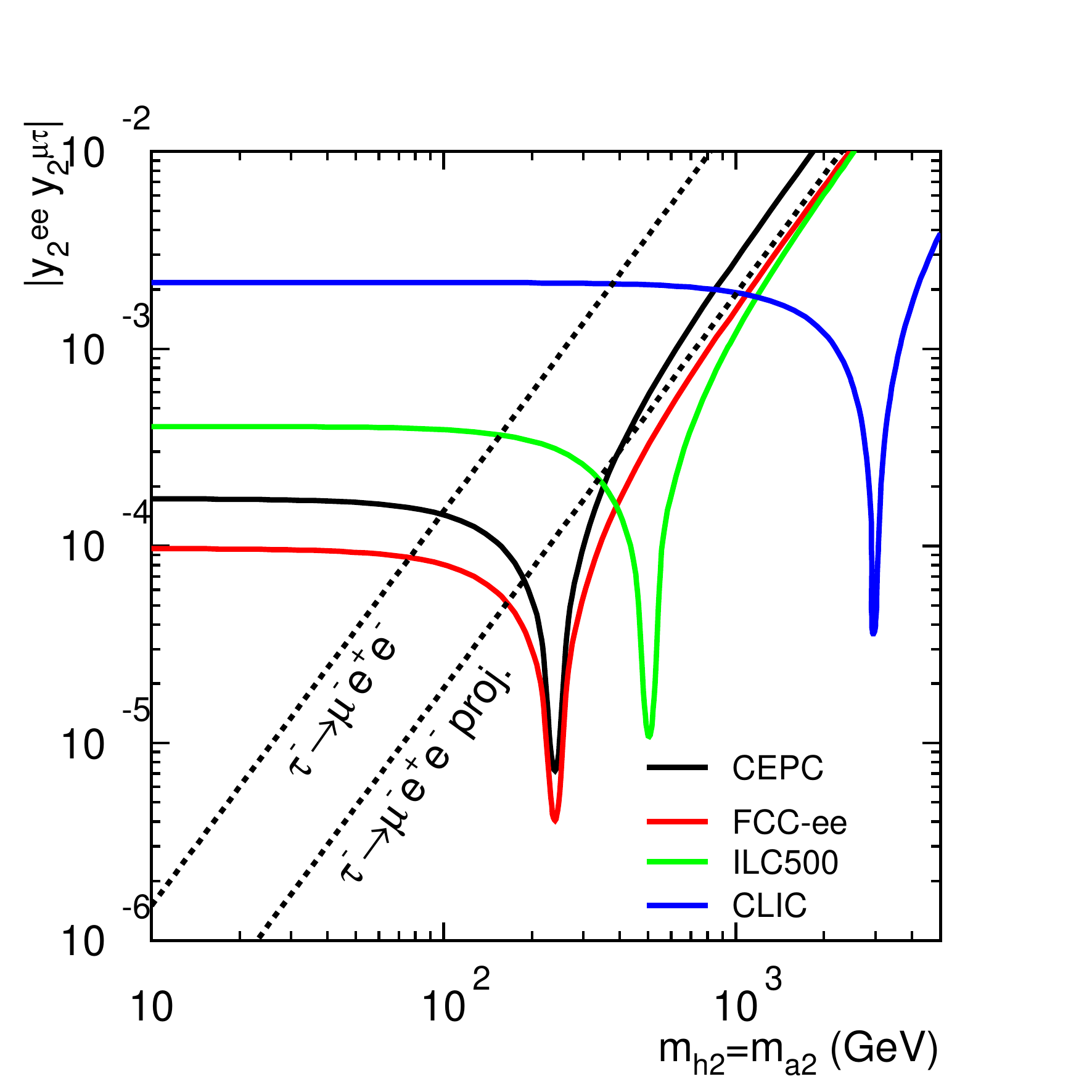}
\includegraphics[scale=1,width=6.5cm]{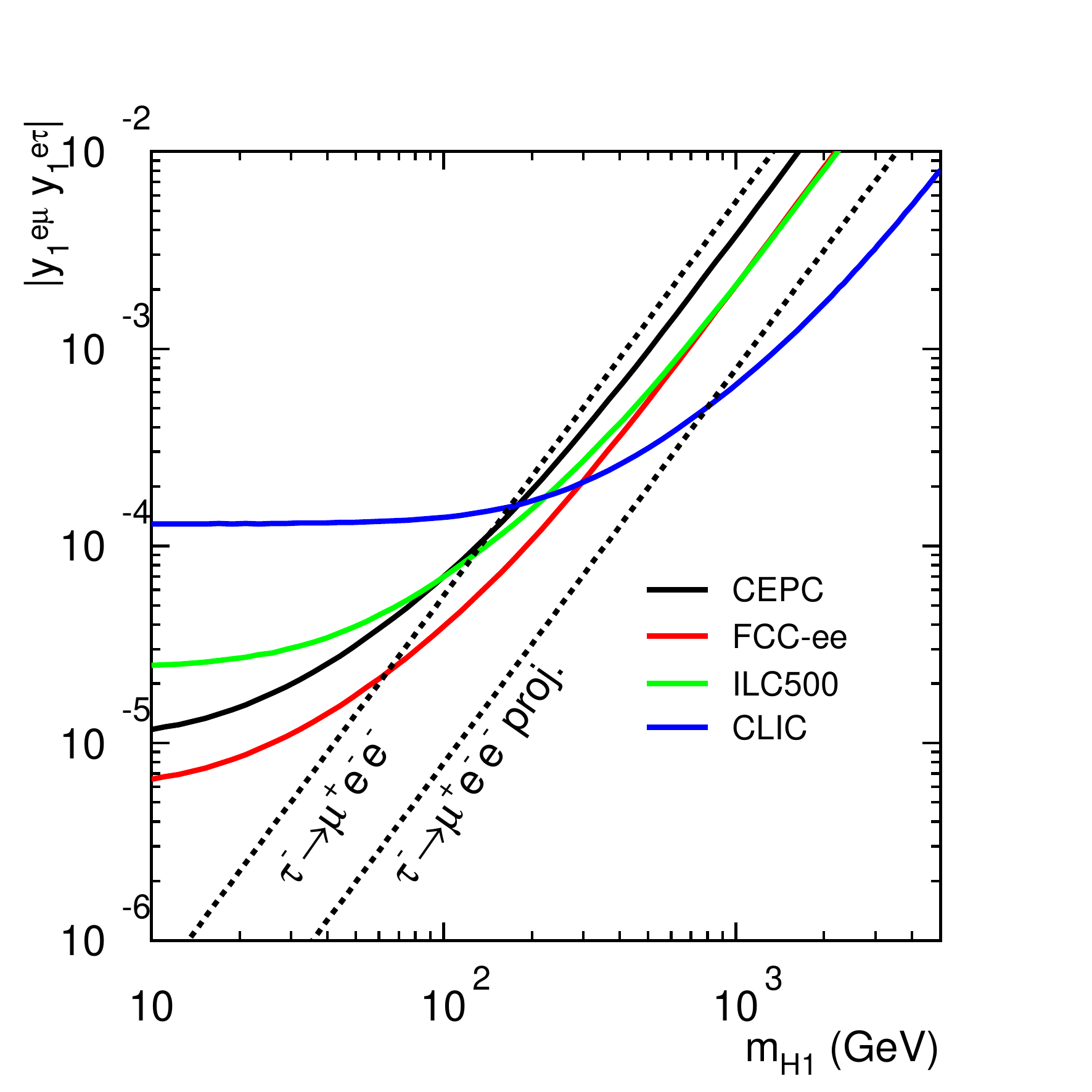}
\includegraphics[scale=1,width=6.5cm]{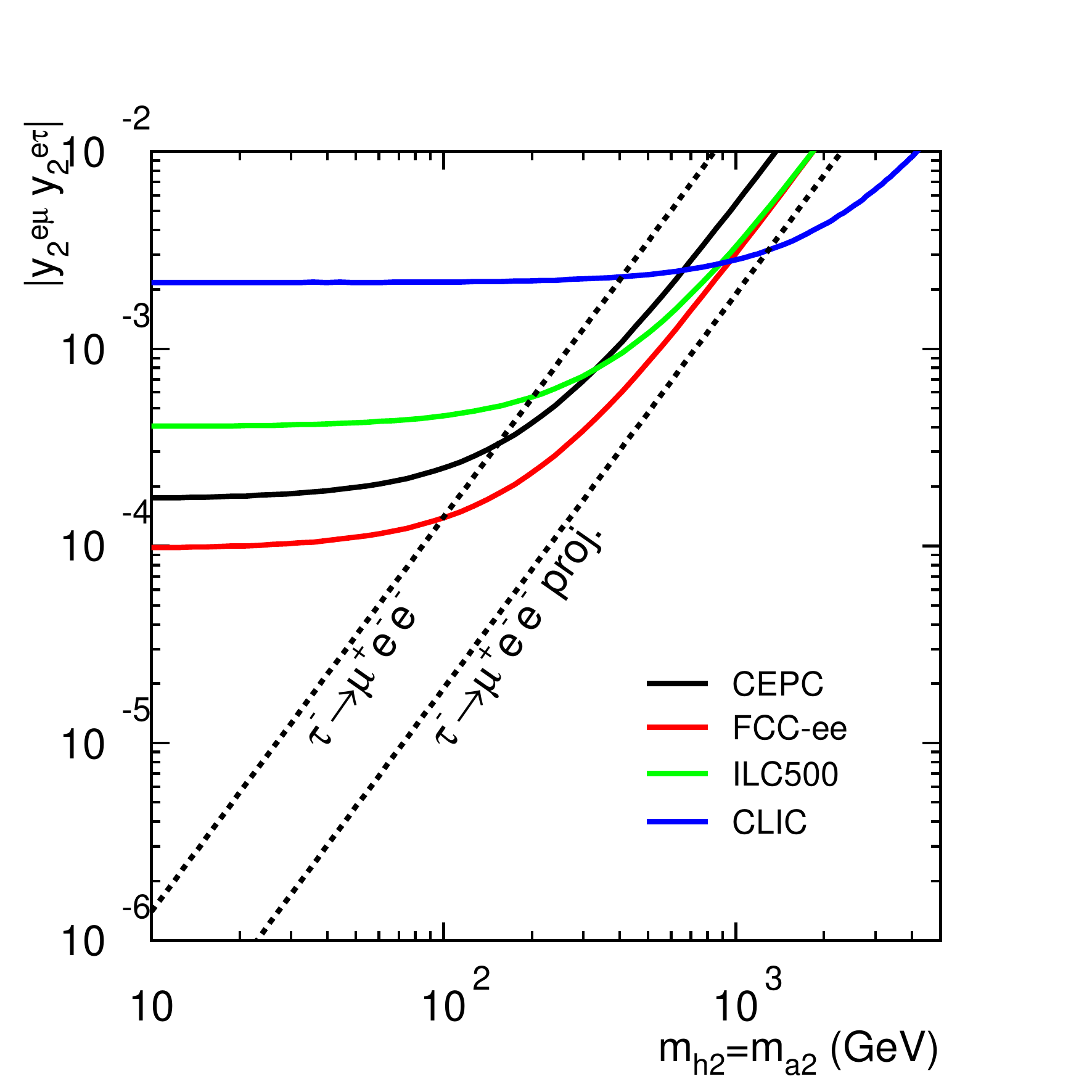}
\caption{Sensitivity to $|y_1^{ee}y_1^{e\mu(e\tau)}|$ (top) through $e^+e^-\to e^\pm\mu^\mp(e^\pm\tau^\mp)$, and $|y_1^{ee}y_1^{\mu\tau}|$ (center) and $|y_1^{e\mu}y_1^{e\tau}|$ (bottom) through $e^+e^-\to \mu^\pm\tau^\mp$, as a function of the mediator mass for $H_1^0$ (left) and $H_2^0$ (right) interaction. The tau efficiency is not included and thus the sensitivity becomes weaker by $77\%$ if there is one tau lepton.}
\label{H1H2}
\end{figure}

In the left panels of Fig.~\ref{H1H2}, we show the lepton collider sensitivity to the detection of
$|y_1^{ee}y_1^{e\mu(e\tau)}|$ (top left) through $e^+e^-\to
e^\pm\mu^\mp(e^\pm\tau^\mp)$, and $|y_1^{ee}y_1^{\mu\tau}|$ (center left) and
$|y_1^{e\mu}y_1^{e\tau}|$ (bottom left) through $e^+e^-\to \mu^\pm\tau^\mp$,
as a function of $m_{H_1}$ for the $H_1^0$ interactions. The reachable limits for the couplings in $H_2^0$ interaction are given in the right panels of
Fig.~\ref{H1H2}, assuming $m_{h_2}=m_{a_2}$.
As stated before,
the $e$ and $\mu$ labels of the couplings in these and following figures should
be swapped to obtain the couplings which can be probed at a muon collider.

For processes with an s channel contribution in the top four figures, there is an
increased sensitivity at the c.m.~energy of the lepton collider due to
resonant enhancement. The scattering through both s and t
channels can be sensitive in both low and high mass region, in particular for
$H_1^0$, due to the interference of two channels. Above the c.m.~energy there are only contributions from
off-shell processes and thus the scattering cross section and consequently the limits on the Yukawa couplings weaken with increasing mass like $\max (y) \propto m$.
The tree-level mediated LFV decays provide a stronger
constraint than the loop-induced radiative decays and existing collider constraints. Thus, in these plots and below, we only include the most stringent constraints from tree-level LFV decays for each relevant coupling and their future projections. A comparison with the
sensitivity of a future $e^+ e^-$ lepton collider demonstrates the
complementarity of the two searches. Low-energy precision experiments generally
dominate for $e\mu$ final states, while a lepton collider provides complementary sensitivity for final states with $\tau$ lepton(s), in particular close to an s-channel resonance $\sqrt{s}\simeq m$.
\begin{figure}[tbp!]
\begin{center}
\includegraphics[scale=1,width=6.5cm]{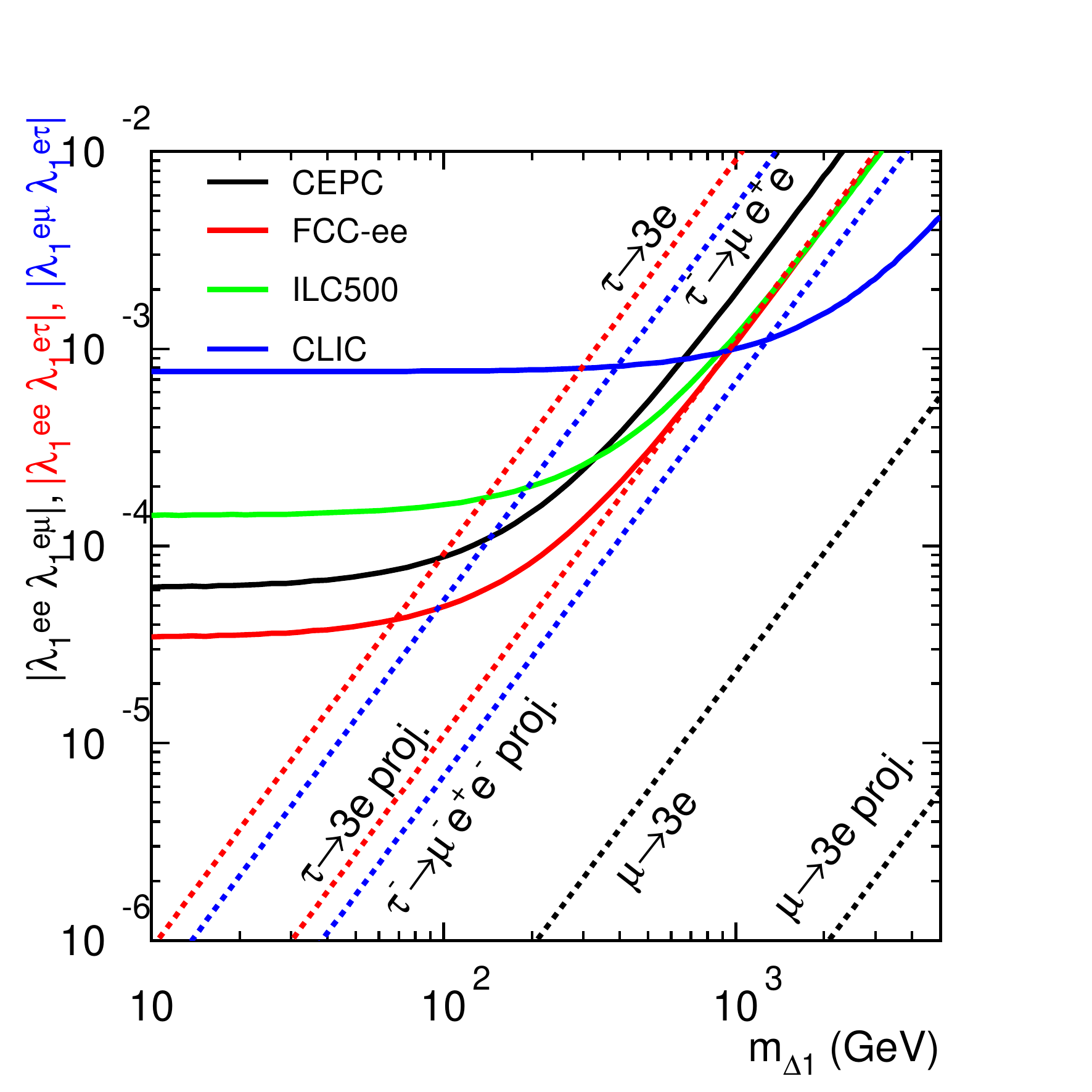}
\includegraphics[scale=1,width=6.5cm]{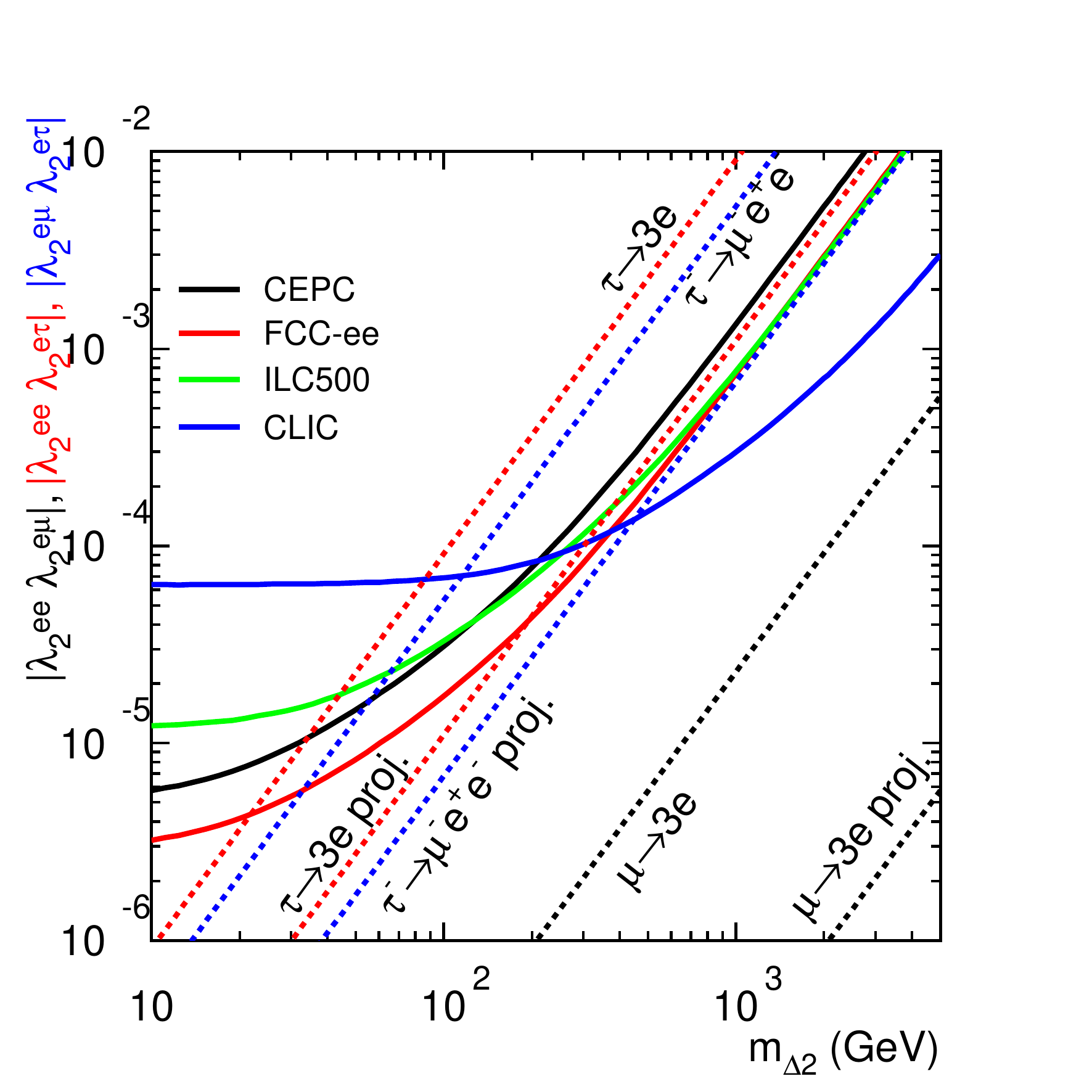}
\end{center}
\caption{Left: Sensitivity to $|\lambda_1^{ee}\lambda_1^{e\mu(e\tau)}|$ through $e^+e^-\to e^\pm\mu^\mp(e^\pm\tau^\mp)$, and $|\lambda_1^{e\mu}\lambda_1^{e\tau}|$ through $e^+e^-\to \mu^\pm\tau^\mp$, as a function of $m_{\Delta_1}$ for $\Delta_1^{++}$ interaction. Right: Sensitivity to $|\lambda_2^{ee}\lambda_2^{e\mu(e\tau)}|$ through $e^+e^-\to e^\pm\mu^\mp(e^\pm\tau^\mp)$, and $|\lambda_2^{e\mu}\lambda_2^{e\tau}|$ through $e^+e^-\to \mu^\pm\tau^\mp$, as a function of $m_{\Delta_2}$ for $\Delta_2^{++}$ interaction.
}
\label{D12}
\end{figure}

The left panel of Fig.~\ref{D12} shows the sensitivity to
$|\lambda_1^{ee}\lambda_1^{e\mu(e\tau)}|$ through $e^+e^-\to
e^\pm\mu^\mp(e^\pm\tau^\mp)$, and $|\lambda_1^{e\mu}\lambda_1^{e\tau}|$ through
$e^+e^-\to \mu^\pm\tau^\mp$, as a function of $m_{\Delta_1}$ for
$\Delta_1^{++}$ interaction. They are all induced by the exchange of
$\Delta_1^{++}$ in t channel. The results for $\Delta_2^{++}$ couplings are in
the right panel of Fig.~\ref{D12}. The lepton colliders are more sensitive
to the couplings of $\Delta_2^{++}$ as spin-1 mediator than those with a scalar mediator
$\Delta_1^{++}$. Similarly to the exchange of $\Delta L=0$ mediators, $\mu\to 3e$ decays outperform the sensitivity of a lepton collider. Lepton colliders, however, provide competitive complementary limits, if there are $\tau$ leptons in the final states.

In summary, the $|y_{1(2)}^{ee}y_{1(2)}^{e\mu}|$ and
$|\lambda_{1(2)}^{ee}\lambda_{1(2)}^{e\mu}|$ couplings are highly constrained
by the tree-level decay $\mu\to 3e$ and hence the accessible regions for lepton colliders have already been excluded. In contrast, CLFV couplings
involving $\tau$ flavor are less constrained and a larger parameter space can be probed by future lepton colliders.
Comparing
different machines, the FCC-ee with the highest integrated luminosity in
proposal can explore smaller couplings in the low mass region. In the high mass region, however,
the ILC500 and CLIC with larger c.m.~energies are able to reach a broader
parameter space.

\subsection{Same-sign lepton collision}

We show the CLFV channels and the explored couplings in either $\Delta L=0$ or $\Delta L=2$ Lagrangian at a same-sign lepton collider in Table~\ref{tab:same-sign}. When same-sign leptons collide, the CLFV process led by $\Delta L=0$ and $\Delta L=2$ interactions only emerges in t and s channel, respectively. This type of collider can also probe identical lepton final states, such as $\mu^-\mu^-, \tau^-\tau^-$, which are forbidden in opposite-sign lepton collisions. As a result, single couplings such as $y_{1(2)}^{e\mu}$ or $y_{1(2)}^{e\tau}$ can be measured through a t channel process for $\Delta L=0$ interactions.

\begin{figure}[htbp!]
\begin{center}
\includegraphics[scale=1,width=6.5cm]{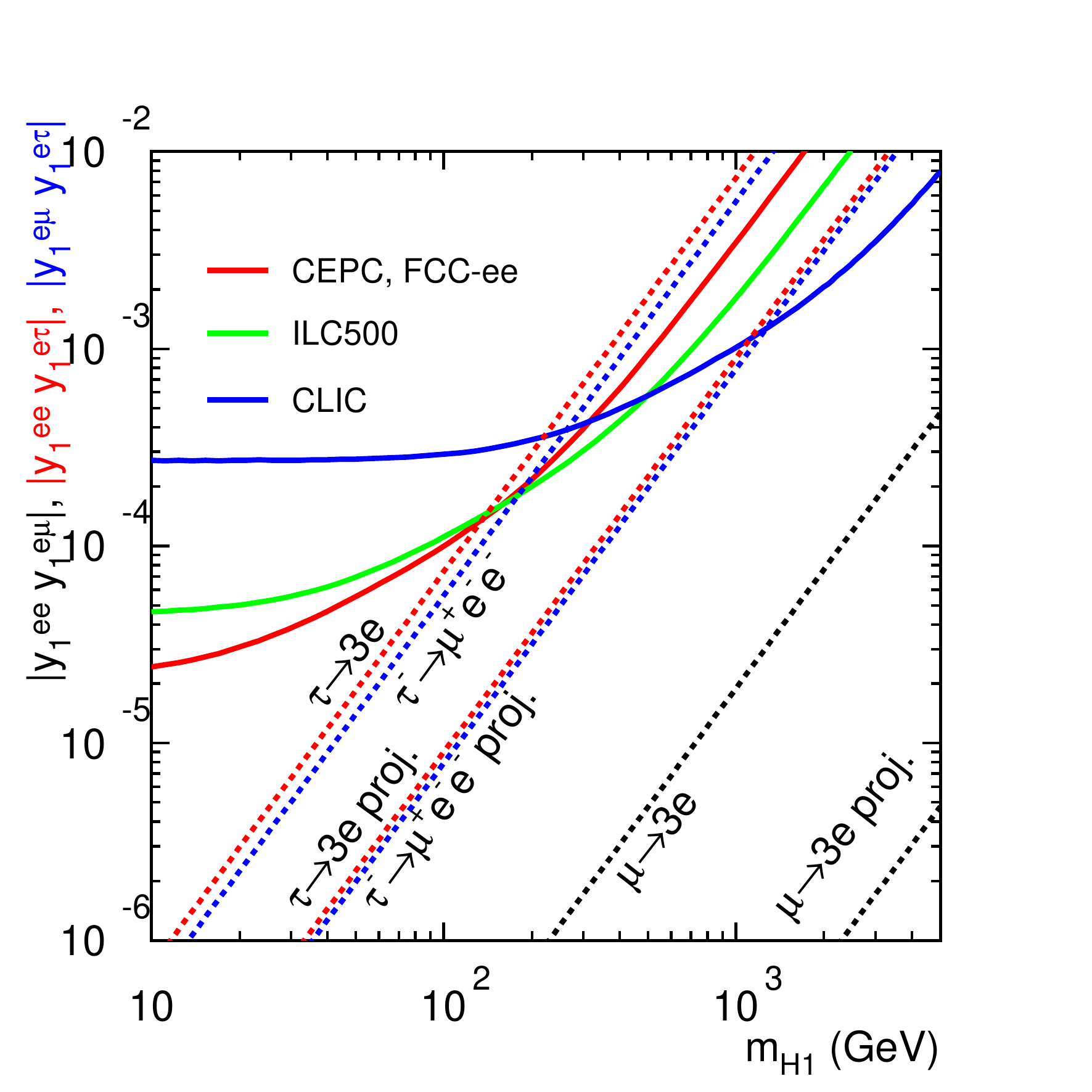}
\includegraphics[scale=1,width=6.5cm]{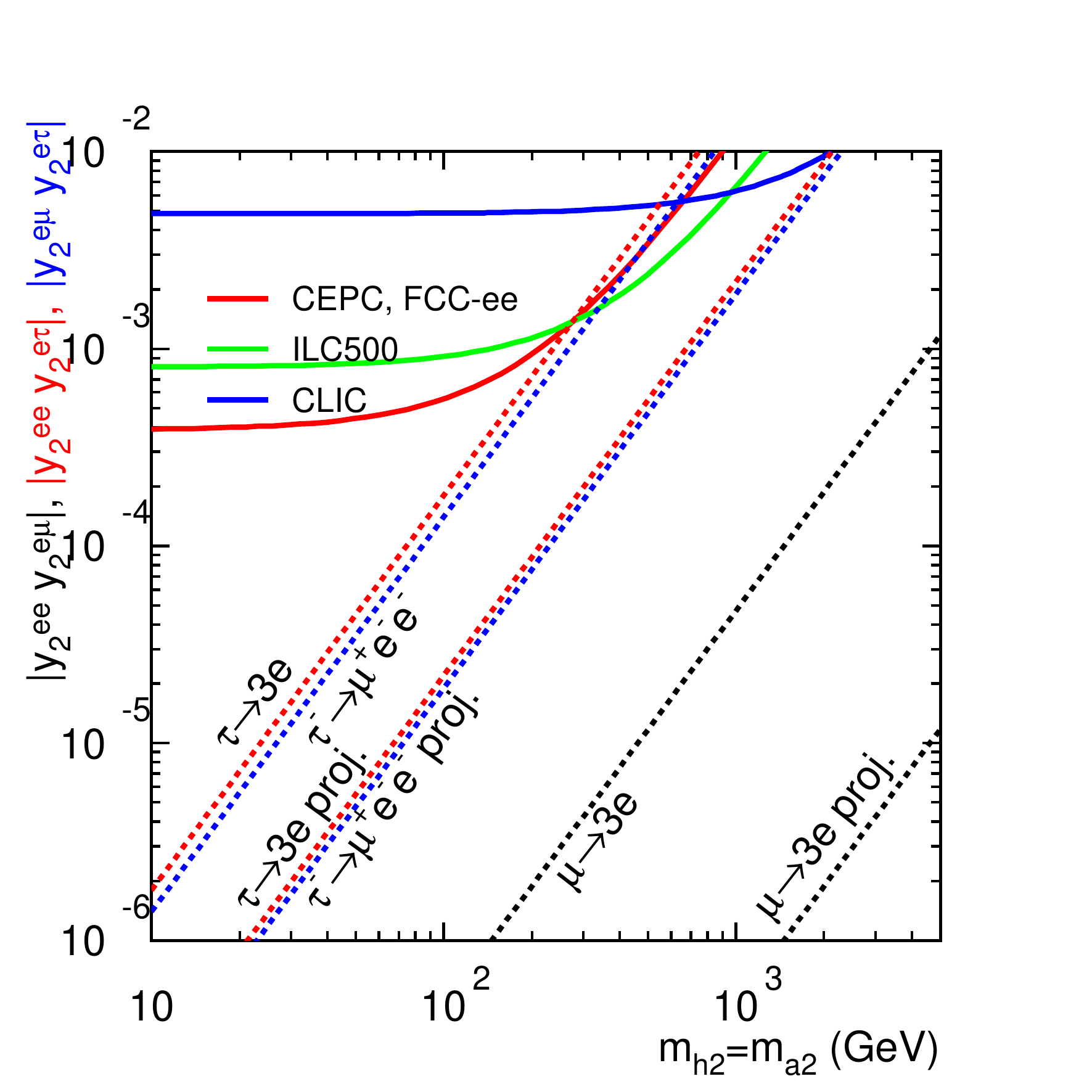}
\end{center}
\caption{Left: Sensitivity to $|y_1^{ee}y_1^{e\mu(e\tau)}|$ through $e^-e^-\to e^-\mu^-(e^-\tau^-)$, and $|y_1^{e\mu}y_1^{e\tau}|$ through $e^-e^-\to \mu^-\tau^-$, as a function of $m_{H_1}$ for $H_1^0$ interaction. Right: Sensitivity to $|y_2^{ee}y_2^{e\mu(e\tau)}|$ through $e^-e^-\to e^-\mu^-(e^-\tau^-)$, and $|y_2^{e\mu}y_2^{e\tau}|$ through $e^-e^-\to \mu^-\tau^-$, as a function of $m_{h_2}=m_{a_2}$ for $H_2^0$ interaction.
}
\label{H12ss}
\end{figure}

\begin{figure}[htbp!]
\begin{center}
\includegraphics[scale=1,width=6.5cm]{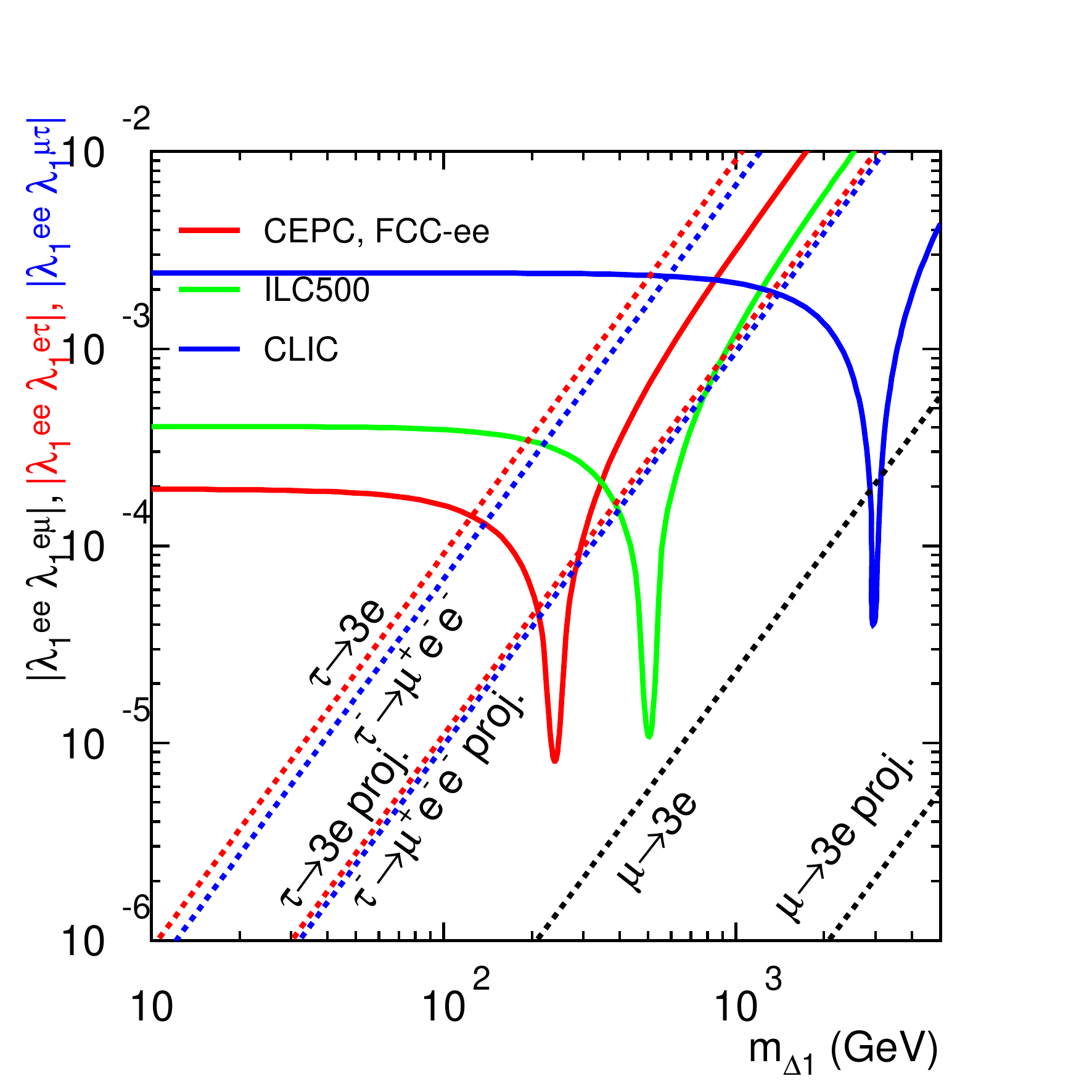}
\includegraphics[scale=1,width=6.5cm]{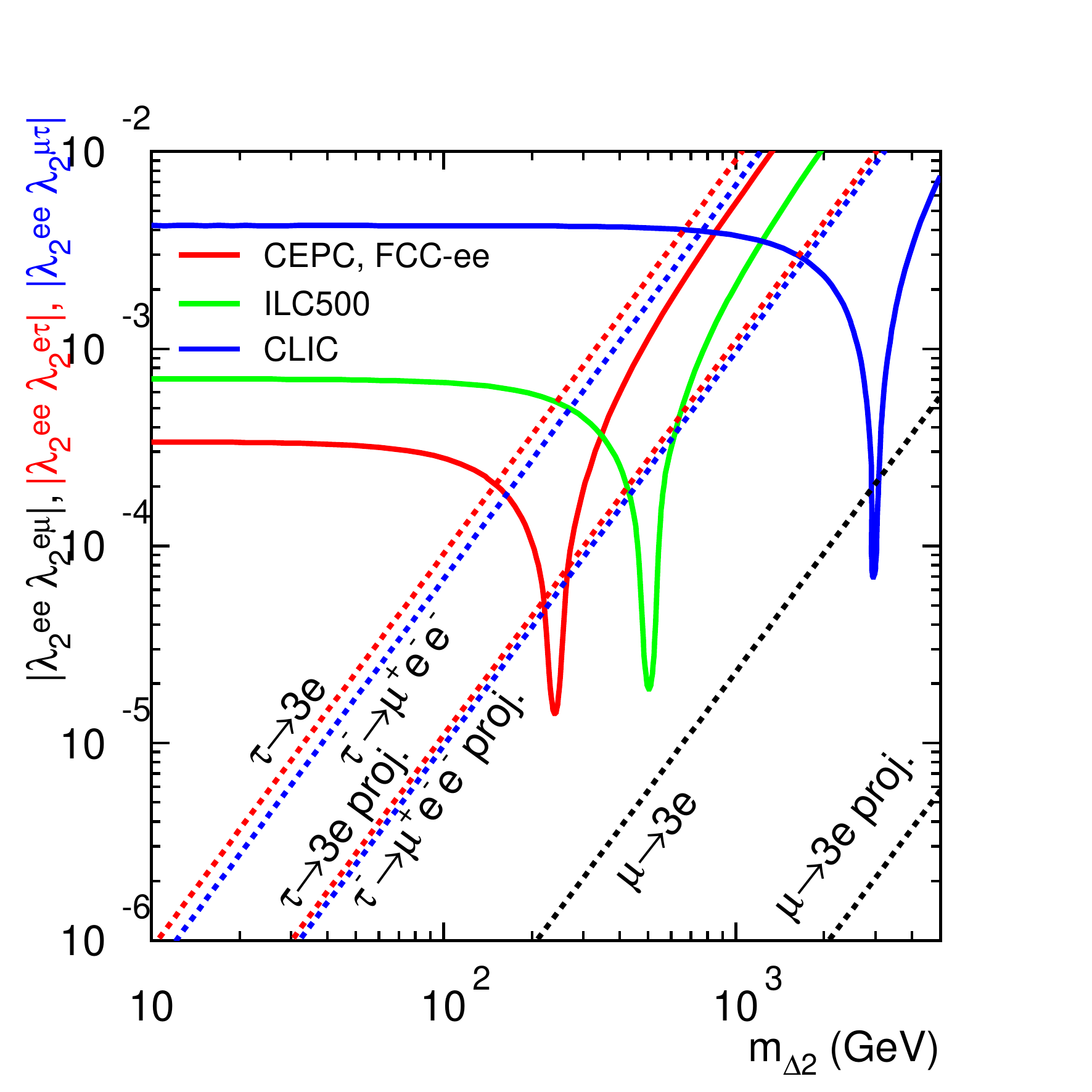}
\end{center}
\caption{Left: Sensitivity to $|\lambda_1^{ee}\lambda_1^{e\mu(e\tau)}|$ through $e^-e^-\to e^-\mu^-(e^-\tau^-)$, and $|\lambda_1^{ee}\lambda_1^{\mu\tau}|$ through $e^-e^-\to \mu^-\tau^-$, as a function of $m_{\Delta_1}$ for $\Delta_1^{++}$ interaction. Right: Sensitivity to $|\lambda_2^{ee}\lambda_2^{e\mu(e\tau)}|$ through $e^-e^-\to e^-\mu^-(e^-\tau^-)$, and $|\lambda_2^{ee}\lambda_2^{\mu\tau}|$ through $e^-e^-\to \mu^-\tau^-$, as a function of $m_{\Delta_2}$ for $\Delta_2^{++}$ interaction.
}
\label{D12ss}
\end{figure}

As the luminosity for an $e^- e^-$ collider will be much smaller than that of $e^+ e^-$ machines, we assume an integrated luminosity of 500 fb$^{-1}$ below for the same-sign lepton colliders. The left panel of Fig.~\ref{H12ss} displays the sensitivity to $|y_1^{ee}y_1^{e\mu(e\tau)}|$ through $e^-e^-\to e^-\mu^-(e^-\tau^-)$, and $|y_1^{e\mu}y_1^{e\tau}|$ through $e^-e^-\to \mu^-\tau^-$, as a function of $m_{H_1}$ for $H_1^0$ interaction. The sensitivity to $|y_1^{e\mu}y_1^{e\mu}| (|y_1^{e\tau}y_1^{e\tau}|)$ through the process with identical final states, i.e. $e^-e^-\to \mu^-\mu^- (\tau^-\tau^-)$, is weakened by a factor of two.
The right panel is for $H_2^0$ couplings. Smaller couplings of vector $H_1^0$ can be reached in t channel, compared with scalar $H_2^0$. We display the explored couplings for $\Delta_1^{++}$ and $\Delta_2^{++}$ through the same scattering as above, but in s channel, in Fig.~\ref{D12ss}. The sensitivity results for processes with identical final states are weaker by a factor of two.

\begin{table}[htbp!]
\begin{center}
\begin{tabular}{|c|c|c|c|}
        \hline
        CLFV channel & flavor $ij,i'j'$ & $\Delta L=0$ & $\Delta L=2$\\
        \hline
        $e^-e^-\to e^-\mu^-$ & $ee,e\mu$ & t & s \\
        \hline
        $e^-e^-\to e^-\tau^-$ & $ee,e\tau$ & t & s \\
        \hline
        $e^-e^-\to \mu^-\mu^-$ & $ee,\mu\mu$ & - & s \\
        \hline
        $e^-e^-\to \mu^-\tau^-$ & $ee,\mu\tau$ & - & s \\
        \hline
        $e^-e^-\to \tau^-\tau^-$ & $ee,\tau\tau$ & - & s \\
        \hline
        $e^-e^-\to \mu^-\mu^-$ & $e\mu,e\mu$ & t & - \\
        \hline
        $e^-e^-\to \mu^-\tau^-$ & $e\mu,e\tau$ & t & - \\
        \hline
        $e^-e^-\to \tau^-\tau^-$ & $e\tau,e\tau$ & t & - \\
        \hline
\end{tabular}
\end{center}
\caption{CLFV channels for probing coupling $y^{ij}y^{i'j'}$ or $\lambda^{ij}\lambda^{i'j'}$ via $\Delta L=0$ or $\Delta L=2$ interactions at $e^-e^-$ collider, while $e\leftrightarrow\mu$ for $\mu^-\mu^-$ collider.}
\label{tab:same-sign}
\end{table}

\section{Conclusion}
\label{sec:Con}

We perform a comprehensive study of the sensitivity to both opposite- and same-sign lepton colliders to charged lepton flavor violating interactions. We consider the most general Lagrangian coupling of two charged leptons to new bosonic particles, involving both $\Delta L=0$ and $\Delta L=2$ interactions. The CLFV processes are mediated by off-shell new particles at tree-level in the scattering processes. Low-energy precision experiments and future lepton colliders provide complementary limits on CLFV couplings. Although low-energy precision experiments provide very stringent limits on the LFV transition $\mu\to e$, the CLFV couplings involving $\tau$ flavor are less constrained.
We find that a large space of these couplings can be probed by future lepton colliders. In general, the FCC-ee with the highest integrated luminosity in proposal can reach smaller couplings in the low mass region. The ILC500 and CLIC with larger c.m.~energies are able to explore broader parameter space in the high mass region.

\acknowledgments
MS thanks Julian Berengut for discussions. Feynman diagrams were created with Ti\textit{k}Z-Feynhand~\cite{Dohse:2018vqo}. This work has been supported in part by the Australian Research Council.

\appendix

\section{Technical details for current constraints}

In this appendix we summarize the analytic expressions for several relevant low-energy constraints and the relevant effective operators for the LEP limits. We generally compare the CLFV process to the related SM process with partial width $\Gamma(\ell\to\ell'\nu\bar\nu')=G_F^2 M^5/192\pi^3$ assuming any new physics contribution can be neglected. $M$ denotes the mass of the decaying lepton $\ell$.
The relevant SM branching ratios are BR($\mu\to e \nu_\mu\bar\nu_e)\approx 1$, BR$(\tau\to e \nu_\tau \bar \nu_e)\approx 0.178$, and BR$(\tau\to \mu \nu_\tau\bar\nu_\mu)\approx 0.174$~\cite{Tanabashi:2018oca}. In the following we report the leading order expressions, where we neglect final state lepton masses and only show the leading order term in the expansion of the internal fermion mass versus the mass of the new boson.

Anomalous magnetic moments only provide mild constraints for the parameter choices in the main part of the text, unless the amplitude is enhanced by a heavy fermion in the loop. There is a $>3\sigma$ discrepancy between the SM prediction, $a_\mu^{SM}=(11659182.3\pm4.3) \times 10^{-10}$~\cite{Tanabashi:2018oca} of the muon anomalous magnetic moment and its measured value, $a_\mu^{exp}=(11659208.9\pm6.3)\times 10^{-10}$~\cite{Tanabashi:2018oca}. The different errors have been added in quadrature. Similarly there is a $2.5\sigma$ discrepancy~\cite{Parker:2018vye} between the SM prediction of the anomalous magnetic moment of the electron, $a_e^{SM}=(1159652181.61\pm0.23)\times 10^{-12}$, and the experimental value, $a_e^{exp}=(1159652180.91\pm 0.26)\times 10^{-12}$~\cite{Tanabashi:2018oca}. In order to derive a constraint we add experimental and theoretical errors in quadrature and demand that the new physics contribution deviates from the experimental observation by at most $3\sigma$ for the electron and $4\sigma$ for the muon in order to take into account the current discrepancies between experiment and SM theory prediction.

\subsection{$H_1^{(\prime)}$}
The branching ratio of the trilepton decay $\ell_0^- \to \ell_1^- \ell_2^+ \ell_3^-$ in the limit of vanishing final state masses is given by
\begin{equation}
\frac{BR(\ell_0\to \ell_1\bar\ell_2\ell_3)}{BR(\ell_0\to \ell^\prime \nu \bar\nu^\prime)} =
\frac{|y_1^{(\prime)23} y_1^{(\prime)01}|^2+|y_1^{(\prime)21} y_1^{(\prime)03}|^2 + 4 \mathrm{Re}(y_1^{(\prime)23*}y_1^{(\prime)01*} y_1^{(\prime)21} y_1^{(\prime)03})}{8 (1+\delta_{\ell_1\ell_3}) G_F^2 m_{H_1^{(\prime)}}^4}
\end{equation}
in Feynman gauge. For radiative LFV decays we find
\begin{align}
	\frac{BR(\ell_1\to\ell_2 \gamma)}{BR(\ell_1\to \ell_2\nu_1\bar\nu_2)}
	= \frac{\alpha_{\rm em} |(y_1^{(\prime)\dagger} y_1^{(\prime)})_{21}|^2  }{12\pi G_F^2 m_{H_1^{(\prime)}}^{4}}
\end{align}
in the limit of vanishing final state masses and neglecting the fermion mass in the loop compared the gauge boson mass. Similarly there is a contribution to the anomalous magnetic moment of the lepton $\ell$
\begin{equation}
	\Delta a_{\ell} = \frac{(y_1^{(\prime)\dagger} y_1^{(\prime)})_{\ell\ell}}{12\pi^2} \frac{m_\ell^2}{m_{H_1}^{(\prime)2}} \geq 0
\end{equation}
and the relevant interaction Lagrangian for muonium-antimuonium conversion is
\begin{equation}
	\mathcal{L} = \frac{|y_1^{\mu e}|^2}{2m_{H_1}^2} \left[ \bar \mu \gamma^\mu P_L e \right] \left[\bar\mu\gamma_\mu P_L e \right]
	+ \frac{|y_1^{\prime\mu e}|^2}{2m_{H_1^\prime}^{2}} \left[ \bar \mu \gamma^\mu P_R e \right] \left[\bar\mu\gamma_\mu P_R e \right]\;.
\end{equation}
There is a new (t-channel) contribution to $e^+e^-\to \ell^+\ell^-$ scattering cross section from $H_{1\mu}^{(\prime)}$ which is described by the effective operator
\begin{align}
\mathcal{L}_{eff} = \frac{y_1^{(\prime)e\ell}y_1^{(\prime)\ell e}}{(1+\delta_{e\ell}) m_{H_1^{(\prime)}}^2} \bar e \gamma_\mu P_{L(R)} e \bar \ell \gamma^\mu P_{L(R)}\ell
\end{align}
under the assumption that either $y_1^{ee}$ or $y_1^{\ell\ell}$ vanish.

\subsection{$H_2$}
The branching ratio of the trilepton decay $\ell^-_0 \to \ell^-_1 \ell^+_2 \ell^-_3$ mediated by neutral scalar exchange in the limit of vanishing final state masses is given by
\begin{align}
	&\frac{BR(\ell_0\to \ell_1\bar\ell_2\ell_3)}{BR(\ell_0\to \ell_1 \nu \bar\nu_1)} =\frac{1}{128 (1+\delta_{\ell_1\ell_3})}
	\Bigg[
	 \left(\frac{1}{G_F m_{h_2}^2}-\frac{1}{G_F m_{a_2}^2}\right)^2
		\\\nonumber &
\quad	\times	\left(|y_2^{32}y_2^{10}|^2+|y_2^{23}y_2^{01}|^2+|y_2^{12}y_2^{30}|^2+|y_2^{21}y_2^{03}|^2 +
		\mathrm{Re}\left( y_2^{12*} y_2^{32} y_2^{30*} y_2^{10} + y_2^{21}y_2^{23*} y_2^{03} y_2^{01*} \right) \right)
			\\\nonumber & \qquad\qquad
	+	\left(\frac{1}{G_F m_{h_2}^2}+\frac{1}{G_F m_{a_2}^2}\right)^2
		\left(|y_2^{32}y_2^{01}|^2+|y_2^{23}y_2^{10}|^2 + |y_2^{12}y_2^{03}|^2+|y_2^{21}y_2^{30}|^2\right)
		\Bigg]
		\;.
\end{align}
For radiative LFV (taking only the neutral scalar into account) we find
for the branching ratio for $\ell_1\to \ell_2\gamma$
\begin{equation}\label{eq:BrH2}
	\frac{BR(\ell_1\to\ell_2\gamma)}{BR(\ell_1\to \ell_2 \nu_1\bar\nu_2)}
	= \frac{3 \alpha_{em}}{16\pi} \left(\left|\sigma_L^\prime\right|^2 + \left|\sigma_R^\prime\right|^2\right)
\end{equation}
with the fine structure constant $\alpha_{em}$ and
\begin{align}
	\sigma_L^\prime & =
	-\frac{(y_2 y_2^\dagger)^{21}}{12} \left(\frac{1}{G_F m_{h_2}^2}+ \frac{1}{G_F m_{a_2}^2} \right)
	-\sum_k\frac{y_2^{*2k} y_2^{*k1}}{2} \frac{m_k}{m_1} \left(\frac{\ln \tfrac{m_k^2}{m_{h_2}^2} -\tfrac32}{G_F m_{h_2}^2} - \frac{\ln \tfrac{m_k^2}{m_{a_2}^2} -\tfrac32}{G_F m_{a_2}^2} \right)\\
	\sigma_R^\prime & =
	-\frac{(y_2^\dagger y_2)^{21}}{12} \left(\frac{1}{G_F m_{h_2}^2}+ \frac{1}{G_F m_{a_2}^2}  \right)
	-\sum_k\frac{y_2^{k2} y_2^{1k}}{2} \frac{m_k}{m_1} \left(\frac{\ln \tfrac{m_k^2}{m_{h_2}^2} -\tfrac32}{G_F m_{h_2}^2} - \frac{\ln \tfrac{m_k^2}{m_{a_2}^2} -\tfrac32}{ G_F m_{a_2}^2} \right)
\end{align}
in the limit of a vanishing final state mass and to leading order in the fermion mass in the loop.
The anomalous magnetic moment of a lepton $\ell$ receives an additional contribution
\begin{multline}
	\Delta a_\ell =
	-\frac{(y_2^\dagger y_2+y_2y_2^\dagger)^{\ell\ell}}{96\pi^2} \left(\frac{m_\ell^2}{m_{h_2}^2}+ \frac{m_\ell^2}{m_{a_2}^2}  \right)\\
	-\sum_k\mathrm{Re}[y_2^{k\ell}y_2^{\ell k}]  \frac{m_k m_\ell}{8\pi^2} \left(\frac{\ln \tfrac{m_k^2}{m_{h_2}^2} -\tfrac32}{m_{h_2}^2} - \frac{\ln \tfrac{m_k^2}{m_{a_2}^2} -\tfrac32}{m_{a_2}^2} \right)\;.
\end{multline}
Thus, for real symmetric Yukawa couplings and $m_{h_2}=m_{a_2}$, the contribution is always negative. The relevant interaction Lagrangian for muonium-antimuonium conversion is
\begin{multline}
	\mathcal{L} = \frac{1}{16m_{h_2}^2} \left[ \left( y_2^{\mu e} + y_2^{e \mu *} \right)\bar \mu e
+\left( y_2^{\mu e} - y_2^{e \mu *} \right)\bar \mu\gamma_5 e \right]^2\\
- \frac{1}{16m_{a_2}^2} \left[ \left( y_2^{\mu e} - y_2^{e \mu *} \right)\bar \mu e
+\left( y_2^{\mu e} + y_2^{e\mu  *} \right)\bar \mu\gamma_5 e \right]^2
\end{multline}
which reduces in the limit $m_{h_2}=m_{a_2}$ for real and symmetric couplings to
\begin{equation}
	\mathcal{L}
=-\frac{ (y_2^{e\mu})^2}{2 m_{h_2}^2} [\bar \mu \gamma^\mu P_L e] [\bar\mu \gamma_\mu P_R e]\;.
\end{equation}
There is a new (t-channel) contribution to $e^+e^-\to \ell^+\ell^-$ scattering cross section from $H_{1\mu}^{(\prime)}$. In the limit of real symmetric Yukawa couplings and $m_{h_2}=m_{a_2}$ we obtain 
\begin{align}
\mathcal{L}_{eff} = -\frac{|y_2^{e\ell}|^2}{2(1+\delta_{e\ell})m_{h_2}^2} \left[ \bar e \gamma_\mu P_{L} e \bar \ell \gamma^\mu P_{R}\ell + \bar e \gamma_\mu P_{R} e \bar \ell \gamma^\mu P_{L}\ell \right]
\end{align}
under the assumption that either $y_2^{ee}$ or $y_2^{\ell\ell}$ vanish.

If the neutral complex scalar originates from a two Higgs doublet model, there is a second contribution to radiative LFV decays from the charged scalar and thus the branching ratio is given by Eq.~\eqref{eq:BrH2} by using the following replacements
\begin{align}
	\sigma_L^\prime & \to \sigma_L^\prime +\frac{ (y_2 y_2^\dagger)_{21}}{24G_F m_{h^+}^2}&
	\sigma_R^\prime & \to \sigma_L^\prime +\frac{ (y_2^\dagger y_2)_{21}}{24 G_F m_{h^+}^2}
\end{align}
and the anomalous magnetic moment receives an additional contribution
\begin{equation}
	\Delta a_\ell \to \Delta a_\ell + \frac{ (y_2^\dagger y_2+y_2 y_2^\dagger)_{\ell\ell}}{192\pi^2} \frac{m_\ell^2}{m_{h^+}^2}\;.
\end{equation}
The mass of the charged scalar is denoted $m_{h^+}$.
Note that there is a partial cancellation between the contribution of the charged scalar and the neutral scalars.
Thus, depending on the masses, the contribution to the anomalous magnetic moment may be positive or negative.

\subsection{$\Delta_1^{++}$}
The branching ratio of the trilepton decay $\ell_0^- \to \ell_1^+ \ell_2^- \ell_3^-$ in the limit of vanishing final state masses is given by
\begin{equation}
	\frac{BR(\ell_0\to \bar\ell_1\ell_2\ell_3)}{BR(\ell_0\to \ell_2 \nu_0 \bar\nu_2)} = \frac{|\lambda_1^{01}\lambda_1^{23*}|^2}{2G_F^2 m_{\Delta_1}^4 (1+\delta_{\ell_2\ell_3})}\;.
\end{equation}
For radiative LFV decays decays we find
\begin{align}
	\frac{BR(\ell_1\to\ell_2 \gamma)}{BR(\ell_1\to \ell_2\nu_1\bar\nu_2)} & = \frac{ \alpha_{\rm em} |(\lambda_1^\dagger \lambda_1)_{21}|^2  }{48\pi G_F^2 m_{\Delta_1}^4}
\end{align}
and the anomalous magnetic moment is given by
\begin{equation}
	\Delta a_{\ell} = \frac{(\lambda_1^\dagger \lambda_1)_{\ell\ell}} {24\pi^2} \frac{m_\ell^2}{m_{\Delta_1}^{2}} \geq 0\;.
\end{equation}
The relevant interaction Lagrangian for muonium-antimuonium conversion is
\begin{equation}
	\mathcal{L} = \frac{\lambda_1^{ee} \lambda_1^{\mu\mu*}}{2m_{\Delta_1}^2} \left[ \bar \mu \gamma^\mu P_R e \right] \left[\bar\mu\gamma_\mu P_R e \right]\;.
\end{equation}
There is a new (t-channel) contribution to $e^+e^-\to \ell^+\ell^-$ scattering cross section from $\Delta_1^{++}$ which is described by the effective operator
\begin{align}
\mathcal{L}_{eff} = \frac{2|\lambda_1^{e\ell}|^2}{(1+3\delta_{e\ell})m_{\Delta_1}^2} \bar e \gamma_\mu P_R e \bar \ell \gamma^\mu P_R \ell \;.
\end{align}

\subsection{$\Delta_3^{++}$}
We can directly translate the results from $\Delta_1^{++}$. The branching ratio of the trilepton decay $\ell_0^- \to \ell_1^+ \ell_2^- \ell_3^-$ in the limit of vanishing final state masses is given by
\begin{equation}
	\frac{BR(\ell_0\to \bar\ell_1\ell_2\ell_3)}{BR(\ell_0\to \ell_2 \nu_0 \bar\nu_2)} = \frac{|\lambda_3^{01}\lambda_3^{23*}|^2}{2G_F^2 m_{\Delta_3}^4 (1+\delta_{\ell_2\ell_3})}\;.
\end{equation}
For radiative LFV decays we find
\begin{align}
	\frac{BR(\ell_1\to\ell_2 \gamma)}{BR(\ell_1\to \ell_2\nu_1\bar\nu_2)} & = \frac{\alpha_{\rm em} |(\lambda_3^\dagger \lambda_3)_{21}|^2 }{48\pi G_F^2 m_{\Delta_3}^4}
\end{align}
in the limit of vanishing final state masses and fermion masses in the loop.
Similarly there is a contribution to the anomalous magnetic moment of the lepton $\ell$
\begin{equation}
	\Delta a_{\ell} = \frac{(\lambda_3^\dagger \lambda_3)_{\ell\ell}} {24\pi^2} \frac{m_\ell^2}{m_{\Delta_3}^{2}} \geq 0\;.
\end{equation}
The relevant interaction Lagrangian for muonium-antimuonium conversion is
\begin{equation}
	\mathcal{L} = \frac{\lambda_3^{ee} \lambda_3^{\mu\mu*}}{2m_{\Delta_3}^2} \left[ \bar \mu \gamma^\mu P_L e \right] \left[\bar\mu\gamma_\mu P_L e \right]\;.
\end{equation}
There is a new (t-channel) contribution to $e^+e^-\to \ell^+\ell^-$ scattering cross section from $\Delta_3^{++}$ which is described by the effective operator
\begin{align}
\mathcal{L}_{eff} = \frac{2|\lambda_3^{e\ell}|^2}{(1+\delta_{e\ell})m_{\Delta_3}^2} \bar e \gamma_\mu P_L e \bar \ell \gamma^\mu P_L \ell \;.
\end{align}

In the Type II Seesaw model there are additional contributions to the radiative decays and the anomalous magnetic moments due to the additional charged scalars.
In the limit of vanishing final state masses and fermion masses in the loop we find for radiative LFV decays
\begin{align}
	\frac{BR(\ell_1\to\ell_2 \gamma)}{BR(\ell_1\to \ell_2\nu_1\bar\nu_2)} & = \frac{\alpha_{\rm em} |(\lambda_3^\dagger \lambda_3)_{21}|^2 }{48\pi}\left(\frac{1}{G_F m_{\Delta_3}^2} +\frac{1}{2\,G_F m_{\Delta_3^+}^2} \right)^2
\end{align}
and the anomalous magnetic moment is changed by
\begin{equation}
	\Delta a_{\ell} = \frac{(\lambda_3^\dagger \lambda_3)_{\ell\ell}} {24\pi^2} \left(\frac{m_\ell^2}{m_{\Delta_3}^{2}} + \frac{m_\ell^2}{2 m_{\Delta_3^+}^2}\right)\geq 0\;,
\end{equation}
where $m_{\Delta_3^+}$ ($m_{\Delta_3}$) denotes the singly (doubly) charged scalar mass of the electroweak triplet scalar.

\subsection{$\Delta_2^{++}$}
The branching ratio of the trilepton decay $\ell_0^- \to \ell_1^+ \ell_2^- \ell_3^-$ in the limit of vanishing final state masses is given by
\begin{equation}
	\frac{BR(\ell_0\to \bar\ell_1\ell_2\ell_3)}{BR(\ell_0\to \ell_2 \nu_0 \bar\nu_2)} =\frac{ \left(|\lambda_2^{01}|^2+|\lambda_2^{10}|^2\right)\left( |\lambda_2^{23}|^2+|\lambda_2^{32}|^2\right)}{8 G_F^2 m_{\Delta_2}^4 (1+\delta_{\ell_2\ell_3})}\;.
\end{equation}
For radiative LFV decays we find
\begin{align}
	\frac{BR(\ell_1\to\ell_2 \gamma)}{BR(\ell_1\to \ell_2\nu_1\bar\nu_2)} & = \frac{49 \alpha_{\rm em} |(\lambda_2^\dagger \lambda_2)_{21}|^2  }{48\pi\,G_F^2 m_{\Delta_2}^4}
\end{align}
in the limit of vanishing final state masses and neglecting the fermion mass in the loop compared the gauge boson mass.
The contribution to the anomalous magnetic moment of the lepton $\ell$ is
\begin{equation}
	\Delta a_\ell = -\frac{7(\lambda_2^\dagger \lambda_2)_{\ell\ell}}{24\pi^2} \frac{m_\ell^2}{m_{\Delta_2}^2} \leq 0
\end{equation}
and the relevant interaction Lagrangian for muonium-antimuonium conversion is
\begin{equation}
	\mathcal{L} = - \frac{\lambda_2^{ee} \lambda_2^{\mu\mu*}}{m_{\Delta_2}^2} \left[\bar \mu \gamma^\mu P_L e\right] \left[\bar \mu  \gamma_\mu P_R e\right]\;.
\end{equation}
There is a new (t-channel) contribution to $e^+e^-\to \ell^+\ell^-$ scattering cross section from $\Delta_{2\mu}^{++}$ which is described by the effective operator
	\begin{align}
		\mathcal{L}_{eff} &=
	-	\frac{|\lambda^{e\ell}_2|^2}{(1+3\delta_{e\ell})m_{\Delta_2}^2}
	\Big[ (\bar \ell \gamma^\mu P_R \ell)(\bar e \gamma_\mu P_L e)	
	- 2(\bar \ell P_R  \ell)(\bar e  P_L e)
	+ {1\over 2}(\bar \ell \Sigma_{\mu\nu} P_R \ell)(\bar e \Sigma^{\mu\nu} P_L e)
	\nonumber\\&\qquad\qquad\qquad\qquad\qquad
	 + (R\leftrightarrow L) \Big]\;.
	\end{align}
This set of operators is not described by the analysis in Ref.~\cite{Abdallah:2005ph}.

\bibliography{refs}
\end{document}